\documentclass{aastex63}
\hypersetup{
    colorlinks=true,
    linkcolor=blue,
    citecolor=blue,
    filecolor=blue,
    urlcolor=blue
}

\usepackage{amsmath,amssymb}
\usepackage{graphicx}
\usepackage{tabularx}
\defcitealias{Leike2020}{L20}
\defcitealias{Zucker2021}{Z21}
\defcitealias{Grosschedl2021}{G21}
\defcitealias{Swiggum2021}{S21}
\defcitealias{Kounkel2020}{K20}
\defcitealias{Forbes2021}{F21}
\begin{document}

\shorttitle{A 3D View of Orion: I. Barnard's Loop}
\shortauthors{Foley et al.}

\graphicspath{{./}{figures/}}

\title{A 3D View of Orion: I. Barnard's Loop}

\correspondingauthor{Michael Foley}
\email{michael.foley@cfa.harvard.edu}

\author[0000-0002-6747-2745]{Michael M. Foley}
\altaffiliation{NSF Graduate Research Fellow}
\affiliation{Center for Astrophysics $\mid$ Harvard \& Smithsonian, 
60 Garden St., Cambridge, MA, USA 02138}

\author[0000-0003-1312-0477]{Alyssa Goodman}
\affiliation{Center for Astrophysics $\mid$ Harvard \& Smithsonian, 
60 Garden St., Cambridge, MA, USA 02138}

\author[0000-0002-2250-730X]{Catherine Zucker}
\altaffiliation{Hubble Fellow}
\affiliation{Space Telescope Science Institute, 3700 San Martin Drive, Baltimore, MD 21218, USA}

\author[0000-0002-1975-4449]{John C. Forbes}
\affiliation{Center for Computational Astrophysics, Flatiron Institute, 162 5th Avenue, New York, NY, 10010, USA}

\author[0000-0001-8235-2939]{Ralf Konietzka}
\affiliation{Center for Astrophysics $\mid$ Harvard \& Smithsonian, 
60 Garden St., Cambridge, MA, USA 02138}
\affiliation{Ludwig-Maximilians-Universität, Geschwister-Scholl Platz 1, 80539 Munich, Germany}

\author[0000-0001-9201-5995]{Cameren Swiggum}
\affiliation{Department of Astrophysics, University of Vienna, Türkenschanzstrasse 17, 1180 Wien, Austria}

\author[0000-0002-4355-0921]{Jo\~{a}o Alves}
\affiliation{Department of Astrophysics, University of Vienna, Türkenschanzstrasse 17, 1180 Wien, Austria}

\author[0000-0001-8135-6612]{John Bally}
\affiliation{Department of Astrophysical and Planetary Sciences, University of Colorado, Boulder, Colorado 80389, USA}

\author[0000-0002-0294-4465]{Juan D. Soler}
\affiliation{Instituto di Astrofisica e Planetologia Spaziali (IAPS). INAF. Via Fosso del Cavaliere 100, 00133 Roma, Italy}

\author[0000-0002-6895-2804]{Josefa E. Gro${\ss}$schedl}
\affiliation{Department of Astrophysics, University of Vienna, Türkenschanzstrasse 17, 1180 Wien, Austria}

\author[0000-0002-0404-003X]{Shmuel Bialy}
\affiliation{Department of Astronomy, University of Maryland, College Park, MD 20742, USA}

\author[0000-0002-1655-5604]{Michael Y. Grudi\'{c}}
\altaffiliation{Hubble Fellow}
\affiliation{Carnegie Observatories, 813 Santa Barbara St.,
Pasadena, CA 91101, USA}

\author[0000-0002-1640-6772]{Reimar Leike}
\affiliation{Max Planck Institute for Astrophysics, Karl-Schwarzschildstra{\ss}e 1,85748 Garching, Germany}
\affiliation{Ludwig-Maximilians-Universität, Geschwister-Scholl Platz 1, 80539 Munich, Germany}

\author[0000-0001-5246-1624]{Torsten En{\ss}lin}
\affiliation{Max Planck Institute for Astrophysics, Karl-Schwarzschildstra{\ss}e 1,85748 Garching, Germany}
\affiliation{Ludwig-Maximilians-Universität, Geschwister-Scholl Platz 1, 80539 Munich, Germany}

\begin{abstract}

Barnard's Loop is a famous arc of H$\alpha$ emission located in the Orion star-forming region. Here, we provide evidence of a possible formation mechanism for Barnard's Loop and compare our results with recent work suggesting a major feedback event occurred in the region around 6 Myr ago. We present a 3D model of the large-scale Orion region, indicating coherent, radial, 3D expansion of the OBP-Near/Brice\~{n}o-1 (OBP-B1) cluster in the middle of a large dust cavity. The large-scale gas in the region also appears to be expanding from a central point, originally proposed to be Orion X. OBP-B1 appears to serve as another possible center, and we evaluate whether Orion X or OBP-B1 is more likely to be the cause of the expansion. We find that neither cluster served as the single expansion center, but rather a combination of feedback from both likely propelled the expansion. Recent 3D dust maps are used to characterize the 3D topology of the entire region, which shows Barnard's Loop's correspondence with a large dust cavity around the OPB-B1 cluster. The molecular clouds Orion A, Orion B, and Orion $\lambda$ reside on the shell of this cavity. Simple estimates of gravitational effects from both stars and gas indicate that the expansion of this asymmetric cavity likely induced anisotropy in the kinematics of OBP-B1. We conclude that feedback from OBP-B1 has affected the structure of the Orion A, Orion B, and Orion $\lambda$ molecular clouds and may have played a major role in the formation of Barnard's Loop. 

\end{abstract}


\section{Introduction} 
\label{sec:intro}
\subsection{Orion Complex and Barnard's Loop}
The Orion complex is the closest site of current massive star formation, located roughly 400-430 pc from the Sun \citep{Menten2007, Zucker2019, Grosschedl2018}. Spanning over 200 deg$^{2}$ on the sky, the region is home to one of the largest and best-studied associations of O and B stars, the so-called Orion OB1 association \citep{Bally2008, Briceno2008}. \citet{Blaauw1964} first identified four major subgroups in the Orion OB1 association, denoted 1a through 1d. \citet{Blaauw1964} found 56 O and B stars in the full Orion OB1 association, while \citet{Bally2008} calculated that Orion should have formed 30-100 stars more massive than 8 M$_{\odot}$ over the past 12 Myr. \citet{Bally2008} thus concludes that 10-20 supernovae likely exploded in the Orion region over the last 12 Myr. The effects of supernovae are indeed well-observed in the Orion region, especially in the Orion-Eridanus superbubble \citep{Reynolds1979, Soler2018}. It is, therefore, unsurprising that Orion is also home to many well-observed examples of the combined effects of ionizing radiation and stellar winds from newly formed massive stars, such as in the Orion Nebula and around the $\lambda$ Orionis star \citep[see, for example,][]{Cunha1996, Mathieu2008, Pabst2020}. We refer to the collective impacts of stellar radiation and winds, including supernovae, on the surrounding interstellar medium as ``stellar feedback". The subgroups in Orion OB1 are believed to have formed sequentially due to stellar feedback, beginning with the dust-free 1a subgroup and ending with dust-embedded 1d subgroup around the Orion Nebula \citep{Blaauw1964, Elmegreen1977, Kubiak2017}. 

Barnard's Loop is one well-studied example of stellar feedback effects in the Orion region \citep{ODell2011, Ochsendorf2015}. This ``Loop", first discovered in 1889 by W.H. Pickering \citep{Pickering1890} and by E.E. Barnard in 1894 \citep{Barnard1894}, is a large arc of H$\alpha$ emission subtending roughly $14^{\circ}$ on the sky \citep{Reynolds1979, Heiles2000}. At a distance of 400 pc, $14^{\circ}$ corresponds to a physical diameter of roughly 100 pc. Barnard's Loop has classically been associated with extended filaments of H$\alpha$ emission throughout the larger Orion region (see blue features in Fig. \ref{fig:main_fig} and the filament located at $180^{\circ} < l < 200^{\circ}, 30^{\circ} < b < 40^{\circ}$). The H$\alpha$ filaments and shells belong to a system of nested shells known as the Orion-Eridanus superbubble, which is believed to have been created by stellar feedback from the Orion OB1 association \citep{Goudis1982, Burrows1993, Bally2008}. Treating Barnard's Loop and the H$\alpha$ filaments as a single shell yields an HII mass of roughly $8 \times 10^{4}$ M$_{\odot}$ d$^{2}_{400}$ and kinetic energy $> 1.7 \times 10^{50}$ d$^{5/2}_{400}$ ergs, where d$_{400}$ is the distance in units of 400 pc \citep{Reynolds1979, Cowie1979, Burrows1993, Bally2008}. The temperature of Barnard's Loop itself is $T\approx6,000$ K with electron density of $n_{e} \approx 2.0$ cm$^{-3}$, similar to, though somewhat cooler than, other HII regions near the Sun \citep{Heiles2000, ODell2011}. 

The exact origin of Barnard's Loop is still debated nearly 130 years after its discovery. Originally, the Loop was thought to be collisionally ionized due to its high relative motion with respect to the ambient ISM \citep{Menon1958}. Subsequent studies showed that ionization instead is dominated by photoionization from the Orion OB1 association, producing a warm, optically-thin, thermal source \citep{ODell1967, ODell2011, Ochsendorf2015}. The formation of Barnard's Loop is also connected with the Orion OB1 association, though it remains debated whether stellar winds and radiation pressure or past supernovae played the dominant role. Recent work has suggested that Barnard's Loop may be distinct from the larger Orion-Eridanus superbubble and extended $H\alpha$ filaments in the region \citep{Ochsendorf2015, Pon2016, Soler2018}, meaning it is one of a number of nested feedback events in the region. Consequently, recent studies have favored the interpretation that Barnard's Loop was produced by an independent supernova that heavily affected the Orion region \citep[e.g.][hereafter \citetalias{Kounkel2020}]{Ochsendorf2015, Kounkel2020}. Such an event may also have been the cause of coherent expansion observed in stellar groups (\citealt{Zari2019}, \citetalias{Kounkel2020}) and gas throughout the Orion complex \citep[][hereafter \citetalias{Grosschedl2021}]{Grosschedl2021}. However, it is important to recall that many supernovae are believed to have exploded throughout the Orion region in the past 12 Myr \citep{Bally2008}. Supernovae may help replenish both the energy in a bubble associated with Barnard's Loop and nested shells throughout the Orion-Eridanus superbubble \citep{Joubaud2019}. 

\subsection{Orion Stars}

\begin{figure*}
    \centering
    \includegraphics[width=1.0\textwidth]{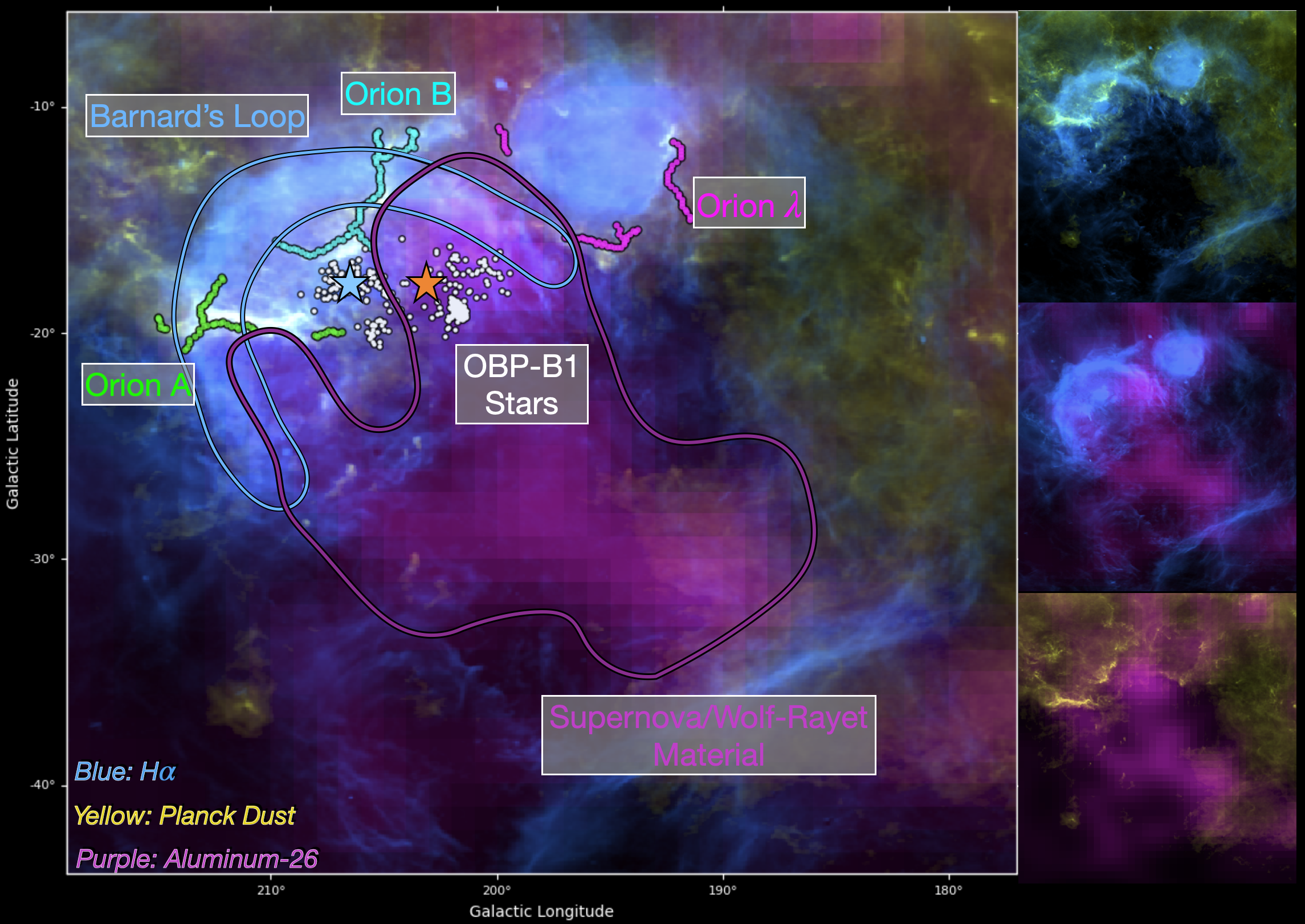}
    \caption{\textit{\textbf{Left:}} Three-color figure of the Orion region. Blue corresponds to H$\alpha$ emission from \citet{Finkbeiner2003}, yellow is extinction measured by \citet{Planck2011}, and purple is $^{26}$Al emission from \citet{Diehl1995}. Spines of the major molecular clouds in the region from \citetalias{Zucker2021} are shown with purple, green, and blue dots corresponding to Orion $\lambda$, Orion A, and Orion B, respectively. White dots correspond to the stars in OBP-B1. The blue star corresponds to the flux center of Barnard's Loop \citep{Reynolds1979}, and the orange star indicates the center of expansion of OBP-B1 from \citetalias{Swiggum2021}. \\
    \textit{\textbf{Right:}} Two-color combinations of the three main observables. \textit{Top right:} H$\alpha$ and extinction; \textit{Middle right:} right displays H$\alpha$ \& $^{26}$Al; \textit{Lower right:} $^{26}$Al and extinction. Subsequent 2D visualizations in this work are different combinations of observables from this figure. 
    \label{fig:main_fig}}
\end{figure*}

Stellar feedback throughout the Orion region is believed to have played an important role in previous and ongoing star formation \citep{Blaauw1964, Elmegreen1977, Lee2009, Ochsendorf2015}. Much of the Orion complex emits brightly in H$\alpha$ and X-rays, typical tracers of stellar feedback (see Figure \ref{fig:main_fig}; \citealt{Haffner2003}, \citetalias{Kounkel2020}). Initially, protostellar jets dominate feedback momentum, but ionizing radiation and stellar winds from massive stars --- especially O-stars --- are responsible for large-scale unbinding and disruption of gas clouds \citep{Dale2013, Rosen2021, Grudic2021_STARFORGE, Grudic2022}. O and B stars are quite common throughout the Orion region, and many of the stars that make up the famous Orion constellation are located within the Orion complex. 

The sizes of shells swept up by winds and radiation depend heavily on the properties of the stars and the ambient conditions of the ISM. Simulations have shown that O-stars are capable of clearing cavities on the order of a $\sim $ few tens of parsecs in radius \citep{Dale2013, ZamoraAviles2019_HII, Grudic2022}. Furthermore, runaway O-stars are known to be quite common, increasing the sphere of influence of stellar feedback from a cluster forming massive stars \citep{Gies1986, Hoogerwerf2001, Gualandris2004, Schoettler2020}. Winds and radiation pressure do much of the early work clearing out gas from the natal environments of clusters with massive stars, enabling the momentum injection from supernovae to be boosted. The combined effect of supernovae, winds, and radiation can produce cavities of hundreds of parsecs in radius \citep{Brown1994_OB1Content, Chevalier1999, Kim2015, Lucas2020}. Such feedback throughout Orion is the cause of the Orion-Eridanus superbubble \citep{Ochsendorf2015, Soler2018}. Considering the presence of the OB1 association and the rich evidence of earlier feedback, the active star formation in Orion is believed to have been (and still is being) triggered by feedback \citep{Bally2008}. 

Ideally, the triggered star formation hypothesis could be tested with information about the 6D-phase space (3 spatial dimensions and 3 velocity dimensions) of both the gas and stars in Orion. Despite Orion's proximity to the Sun, the 6D stellar structure has remained quite uncertain. \textit{Hipparcos} was unable to deliver precise distances to many stars in Orion due to the distance of the complex and unfavorable kinematics \citep{deZeeuw1999}. Working in 2D, a number of studies resolved substructure (both in space and age of stars) in the Orion OB1 association and Orion Nebula Cluster photometrically \citep[e.g.][]{Hillenbrand1997, Jeffries2011, Alves2012, Kubiak2017} or using limited velocity information \citep[e.g.][]{Briceno2007, Furesz2008}. Thanks to Gaia \citep{Gaia2016} and APOGEE \citep{APOGEE_Majewski}, though, 6D stellar information is now available on large scales throughout the Orion region. This has led to rapid constraints on the spatial extents and ages of groups within the Orion OB1 association \citep{ Jerabkova2019, Kos2019, Schoettler2020}. 6D information allows for stellar trajectories to be traced back in time, meaning that hypotheses of disruption by feedback can now be tested more accurately (\citetalias{Kounkel2020}, \citetalias{Grosschedl2021}). 

The hypothesis from \citet{Blaauw1964} of sequential star formation across Orion has largely been supported by the above studies, though the original 10-12 Myr progression from OB1a to OB1d has been significantly augmented by the discovery of spatial and age substructure in the studies mentioned above. For example, \citet{Kos2019} discovered a 21 Myr old stellar population in the Orion complex. \citet{Jerabkova2019} found significant age substructure within the traditional OB1a-d groups, particularly within the OB1a group, along with 17 Myr old spatial substructure in the form of the Orion Relic Filament. Such results suggest that the picture of sequential star formation in Orion isn't as clean as the four step-process proposed in \citet{Blaauw1964}; instead, the large-scale sequential picture may rely on a significant number of individual star-formation events.

Coupling stellar measurements to measurements of gas and dust can provide important dynamical information for understanding the star formation timeline in a region. However, relating stars to the gas and dust observed in 2D remains difficult, primarily due to the challenges of associating gas velocity (from position-position-velocity spectral-line cubes) with a third spatial dimension of distance (in position-position-position space). 3D dust mapping provides a unique alternative. By combining information from 3D dust mapping with maps of dust and gas on the (2D) sky, we can begin to bring structures identified in 2D into 3D space for the first time \citep[][hereafter \citetalias{Zucker2021}]{Zucker2021}.

\subsection{Orion Molecular Clouds and Dust}

Gaseous filamentary or cometary structure is common throughout Orion, further suggesting that feedback has played a substantial role in shaping the current layout of the complex \citep{Bally1987, Bally2008, Gandolfi2008, Bally2010, Smith2020}. In the traditional picture of sequential star formation from \citet{Blaauw1964}, the formation of massive stars in each generation would halt further star formation by evacuating gas and dust from the natal environment \citep{Bally2008, Dale2013, Kim2015, ZamoraAviles2019_HII, Grudic2022}. Evacuated gas would continue expanding in dense shells, sweeping up interstellar material until the shells ran out of energy and merged with the turbulent gas in the interstellar medium. Dense shells may begin collapsing after sweeping up enough material to induce gravitational collapse, serving as the sites for further star formation. Two prominent examples of nearby star formation occurring in dense shells are the PerTau Shell and the Local Bubble \citep{Bialy2021, Zucker2022}. In Orion, feedback sources may have pushed gas from the site of OB1a to the current position of the Orion Nebula and molecular clouds, resulting in the sequential star formation picture. On larger scales, the Orion-Eridanus superbubble is a notable example of star formation promoted by feedback \citep{Lee2009}. 

Recent work has identified another intriguing hypothesis for some of the star formation in Orion: cloud-cloud collisions \citep{Fukui2018, Fujita2021, Enokiya2021}. Cloud-cloud collisions may act in concert with stellar feedback to make Orion an active region of star formation. Recent work by \citet{Skarbinski2022} shows that the majority of cloud-cloud collisions occur at low relative velocities ($<5$ km s$^{-1}$) on the edges of expanding bubbles of hot, ionised gas or through galactic rotation. Such collisions can play a large role in triggering star formation. For example, \citet{Skarbinski2022} find that their simulated galaxy receives a 25\% increase in instantaneous star formation rate with a merger rate of 1\% of clouds per Myr vs. clouds with no mergers. Therefore, while stellar feedback may set the stage for triggered star formation, cloud-cloud collisions could serve as an important mechanism carrying the star formation forward in Orion.   

To what degree is triggered star formation responsible for the star formation observed in Orion? To address this question, it is helpful to use 3D kinematic and spatial data to extrapolate backwards in time. Recent work mapping the 3D spatial distribution of dust in the local solar neighborhood has delivered an unprecedented look at local molecular clouds, including the Orion region (\citealt{RezaeiKh2020}, \citetalias{Zucker2021}, \citealt{Dharmawardena2022}). \citet[][hereafter \citetalias{Leike2020}]{Leike2020} used several catalogs of stellar colors and Gaia positions to map the 3D distribution of dust in the local solar neighborhood, covering much of the Orion region with exquisite parsec-scale resolution. \citetalias{Zucker2021} used the \citetalias{Leike2020} map to create 3D ``skeletons" for many local molecular clouds, including the three major molecular clouds found in Orion: Orion A, Orion B, and Orion $\lambda$. Orion A, Orion B, and Orion $\lambda$ are very well-studied regions of star formation \citep[e.g.][]{Maddalena1986, Dame2001, Mathieu2008, Lombardi2011}. By combining the 3D cloud topologies, the full \citetalias{Leike2020} map, measurements of 3D stellar motions from Gaia, and tracers of dense gas throughout the Orion complex, we can now build an unprecedented 3D view of the Orion region that includes both dust and stars. 

\subsection{OBP-Near and Briceño-1}

\begin{figure*}[ht!]
    \centering
    \includegraphics[width=0.7\textwidth]{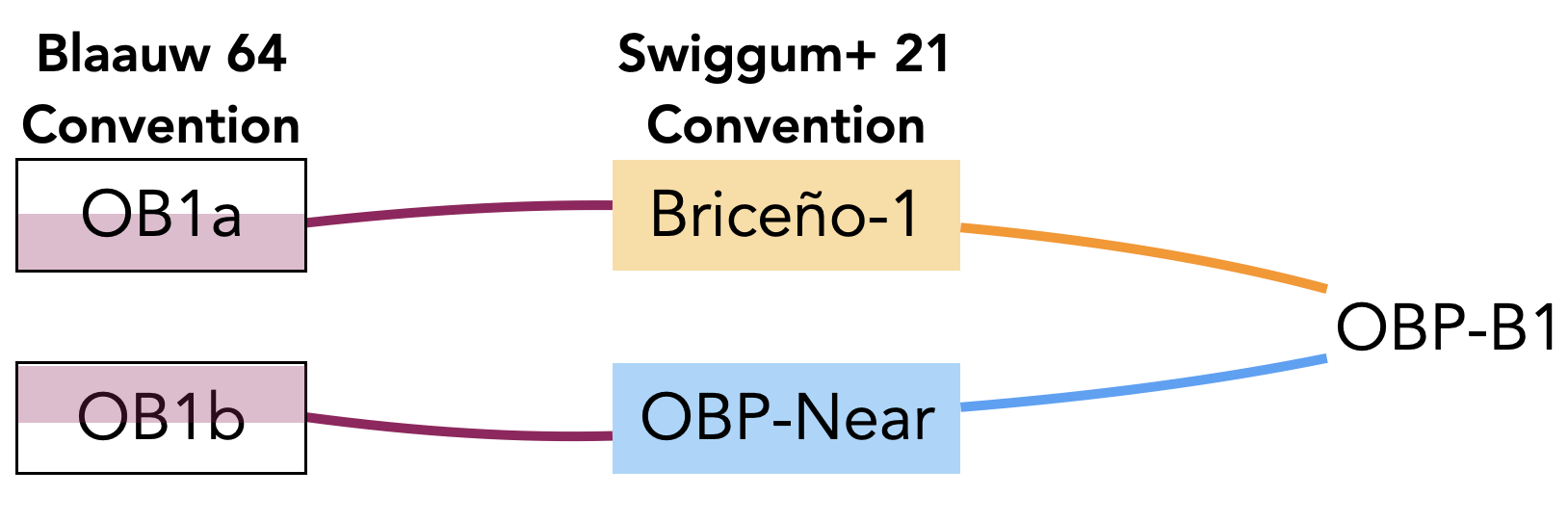}
    \caption{Diagram showing the composition of OBP-B1 from the two different naming conventions: \citet{Blaauw1964} and \citetalias{Swiggum2021}. OBP-B1 is the full combination of Briceño-1 and OBP-Near, which are subgroups of OB1a and OB1b, respectively.   
    \label{fig:obp_name}}
\end{figure*}

In this work, we use the 3D context to focus particularly on an intriguing group of stars found in the OB1a and 1b subgroups. \citet[][hereafter \citetalias{Swiggum2021}]{Swiggum2021} find that two stellar groups first identified in \citet{Chen2020} -- OBP-Near and Briceño-1 -- exhibit coherent radial expansion from a single point. OBP-Near stands for \textit{Orion Belt Population - Near}, and it is part of the substructure identified around the Orion Belt stars and $\sigma$ Orionis \citep{Walter1997, Alves2012}. Briceño-1 has significant overlap with the 25 Orionis population \citep{Briceno2007}. Using the classification from \citet{Blaauw1964}, Briceño-1 and OBP-Near belong to the OB1a and OB1b subgroups, respectively (see Figure \ref{fig:obp_name} for a schematic).

As indicated in Figure \ref{fig:main_fig}, OBP-Near and Briceño-1 lie near the centers of both Barnard's Loop and a large dust cavity seen in 3D. We refer to this cavity as the Orion Shell, part of which was first observed in \citet{RezaeiKh2020}. Both OBP-Near and Briceño-1 lie in front of the Orion A and B molecular clouds (\citetalias{Zucker2021}). As seen on the Sky (Figure \ref{fig:main_fig}), Orion A, B, and $\lambda$ lie along the rim of the Orion Shell, suggesting that the formation of the Orion Shell and the molecular clouds may be coupled. 

Given the similar ages of the two groups and the coherent expansion, we treat the combination of OBP-Near and Briceño-1 as a single population in this paper: OBP-B1. \citetalias{Swiggum2021} find a combined age of $t_{iso} = 6.8^{+3.6}_{-2.9}$Myr for OBP-B1. While this is useful dynamically, there are reasons to believe that OBP-B1 and Briceño-1 are, in fact, two distinct populations. For example, fitting isochrones to OBP-Near and Briceño-1 yields slightly disparate ages. \citetalias{Swiggum2021} find that OBP-Near and Briceño-1 display different disk-fractions, suggesting that OBP-Near is younger than Briceño-1. Furthermore, the the OB1a and OB1b subgroups (to which Briceño-1 and OBP-Near belong, respectively) are believed to have formed at different times. Given this evidence, we also consider the implications for our results if OBP-Near and Briceño-1 are in fact two distinct populations.

\subsection{Paper Layout}

This is the first paper in a set of two which explores the feedback processes and the 3D structure of the Orion star-forming region, including Barnard's Loop. In this paper (Paper I), we specifically focus on OBP-B1 and the cluster's relationship to Barnard's Loop. We use a number of observables to demonstrate that the OBP-B1 cluster of stars likely produced at least one of the supernovae that heavily contributed to the formation of Barnard's Loop. Furthermore, feedback from OBP-B1 potentially helped trigger star formation in adjacent molecular clouds. In the second paper, though, we will argue that many more supernovae beyond OBP-B1 were necessary to completely form Barnard's Loop and define the Orion region.

This paper is organized as follows: Section \ref{sec: data} describes the observational data used for this analysis and a finding chart for the region; Section \ref{sec: 3D_structures} discusses the structures we identify in our 3D map of the region; Section \ref{sec: Loop_conditions} presents further background on the physical conditions of Barnard's Loop; Section \ref{sec: 3D_kinematics} discusses the 3D stellar and gas kinematics in the region; Section \ref{sec: possible_supernovae} presents calculations addressing the possibility of supernovae from OBP-B1; Section \ref{sec: timeline} discusses our timeline for the major events from OBP-B1; Section \ref{sec: barnard_loop_formation} connects the timeline to Barnard's Loop; and our findings are briefly summarized in Section \ref{sec: conclusions}.

\begin{figure*}
    \centering
    \includegraphics[width=1.0\textwidth]{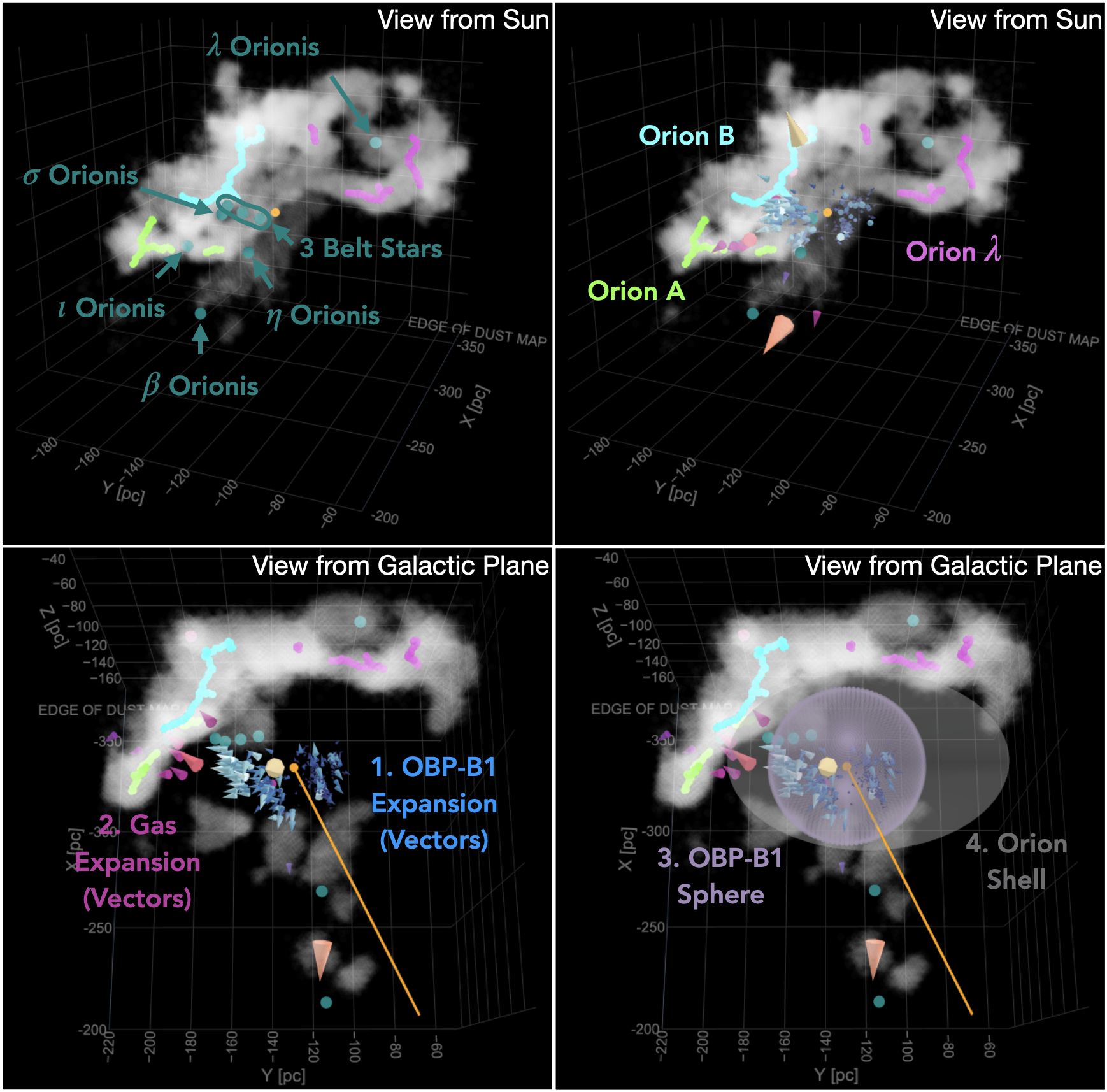}
    \caption{\textit{Top Left:} View from the Sun of the 3D \citetalias{Leike2020} dust map (white) and molecular cloud spines identified in the 3D dust from \citetalias{Zucker2021}. Here, Orion A, B, and $\lambda$ are green, cyan, and pink, respectively. Notable stars within our 3D area of interest (teal) and the expansion center of OBP-B1 (orange) are also shown.\\ 
    \textit{Top Right:} View from the Sun featuring all of the elements of the top left panel as well as OBP-B1 (blue vectors) and gas tracers identified in \citetalias{Grosschedl2021} (purple vectors).\\
    \textit{Bottom Left:} Top-down view (view from the galactic plane) of all previous elements: \citetalias{Leike2020} dust map; Orion A, B, and $\lambda$; notable stars; OBP-B1; gas tracers; and the OBP-B1 expansion center with a line indicating the line-of-sight from the Sun. \\
    \textit{Bottom Right:} View from the galactic plane of all previous elements as well as our models of the OBP-B1 Sphere (purple) and the Orion Shell (gray). 
    An interactive version of the figure can be found \href{https://mfoley-astro.github.io/OrionPaper/orion_region_interactive.html}{here}. Subsequent 3D visualizations in this paper are different rotations or combinations of elements in this figure. 
    \label{fig:3DFindingChart}} 
\end{figure*}

\section{Observational Data}
\label{sec: data}

\subsection{3D Dust Map}
\label{subsec: dust_calc}
The \citetalias{Leike2020} dust map presents high-resolution, 3D data for a large fraction of the local solar neighborhood. In Galactic Cartesian coordinates, their map spans 570 pc in $z$ and 740 pc in both $x$ and $y$. The map is provided in units of opacity density per parsec: $s_x \equiv (\Delta\tau_{\rm G})/(\Delta L/{\rm pc})$. Here, $\tau_{\rm G}$ is the dust opacity in the \textit{Gaia} \textit{G} band, and 
$\Delta\tau_{\rm G}/\Delta L$ is the difference in the \textit{Gaia} \textit{G} band dust opacity per unit length. As in \citet{Bialy2021}, we convert the opacity density to a gas number density ($n$) using the following formula: 
\begin{equation}
\label{eq: n-tau}
    n = 880 \ s_x \ {\rm cm^{-3}} \ .
\end{equation}
This calculation assumes a standard extinction curve from \citet{Draine2011}, which indicates $A_{\rm G}/N = 4\times 10^{-22}$ mag cm$^2$. After applying the conversion factor to every voxel, we use the gas number density for all calculations throughout this paper that reference the \citetalias{Leike2020} map.

Based on the \citetalias{Leike2020} dust map, \citetalias{Zucker2021} derived 3D ``spines'' for many nearby molecular clouds. The spines represent the longest path fit for individual molecular clouds using the 3D dust map. We make use of the spines for Orion A, B, and $\lambda$ in our analysis. It is important to note that the Orion A, B, and $\lambda$ clouds fall near the very edge of the \citetalias{Leike2020} map. Comparison with 2D extinction maps indicates that some of the cloud mass may lie beyond the \citetalias{Leike2020} map boundaries, which is found in \citetalias{Zucker2021} to mildly bias the distance measurements for the spines to be closer than they actually are. However, significant parts of the clouds do fall within the 3D dust map boundaries, and we focus on the cloud sections covered by the 3D map. 

\subsection{YSO Tracers of Gas Kinematics}
\citetalias{Grosschedl2021} presented the first study of the 3D gas dynamics in Orion using proper motions of young stellar objects (YSOs) from Gaia DR2 and CO radial velocities as tracers of gas motion. They argue that the YSOs carry the memory of feedback-driven star-formation in Orion. To study this argument, our work incorporates those gas tracers into our large-scale study of the dynamics in Orion. Vectors indicating gas expansion \citet{Grosschedl2021} come from that work. 

\subsection{Stellar Positions and Kinematics}
The OBP-B1 star cluster was identified in \citetalias{Swiggum2021}. It comprises Brice\~no-1, a group that very closely matches the 25 Orionis group identified in \citet{Briceno2007}, and OBP-Near, a new group discovered near the foreground of the OB1b sub-association \citep{Chen2020}. In the Orion OB1 classification system of \citet{Blaauw1964}, OBP-Near is a subgroup of OB1b and Briceño-1 is a subgroup of OB1a \citep{Briceno2007, Swiggum2021}. Figure \ref{fig:obp_name} shows a diagram of the membership of OBP-B1 from the different naming conventions, and Figure \ref{fig:3DFindingChart} shows all 3D data in an interactive plot. \citetalias{Swiggum2021} relied on data from Gaia DR2 and eDR3, but in this work we use proper motions, radial velocities, and spatial measurements from the full Gaia DR3 release \citep{Gaia_DR3}.

Radial velocities for the stars in OBP-B1 have been obtained from Gaia DR3, Gaia-ESO, GALAH DR3, and APOGEE \citep{Gaia_DR3, Gaia_ESO, APOGEE_Blanton, APOGEE_Majewski, Galah_DR3}. It is important to note that, for subsequent visualizations in this paper featuring the stars of the OBP-B1 cluster, only stars with radial velocity measurements are shown. This represents only $\sim 20\%$ of the total cluster membership. Displaying all the stars extends the stellar distribution but does not appreciably change the center of the cluster. 

\subsection{Notable Massive Stars}
Orion is home to many famous O and B stars. In our 3D Figure \ref{fig:3DFindingChart}, we display all of the notable massive stars in the Orion constellation that fall within our region of interest. Masses of these stars range from 8-40 $M_{\odot}$, depending on the distances assumed. Distances to almost all of the massive stars in Orion are quite uncertain, so the placement of these stars in a 3D context should not be taken as absolute \citep{Van_Leeuwen2007}. Orion's Belt stars in particular are quite uncertain, so we adopt a rough average distance of 380 pc from the literature for all three Belt stars \citep{Brown1994_OB1Content, Van_Leeuwen2007, Hummel2013, Shenar2015, Puebla2016}: $\delta$ Orionis (Mintaka); $\zeta$ Orionis (Alnitak); and $\epsilon$ Orionis (Alnilam).

For $\lambda$ Orionis (Meissa), which is a double-star system composed of an O star and a B star, Gaia DR3 reported separate distances of 386 and 404 pc. Due to distance uncertainties, we average these two distances and display $\lambda$ Orionis as a single point at a distance of 395 pc \citep{Gaia_DR3}. The other stars featured, with their corresponding distances and references, are: $\beta$ Orionis (Rigel), 264 pc \citep{Van_Leeuwen2007}; $\sigma$ Orionis, 387 pc \citep{Schaefer2016-gf}; $\iota$ Orionis, 412 pc \citep{Maiz_Apellaniz2020}; and $\eta$ Orionis, 300 pc \citep{Van_Leeuwen2007}. $\kappa$ Orionis (Saiph), $\gamma$ Orionis (Bellatrix), and $\alpha$ Orionis (Betelgeuse) fall outside our area of focus and are not included. 

\subsection{$^{26}$Al Emission}
$^{26}$Al is a radioactive tracer of supernovae and Wolf-Rayet stars with a half life of $ \tau = 0.7$ Myr. Therefore, it serves as a strong indicator of relatively recent feedback-driving events. We use the full-sky COMPTEL 1.809 MeV $^{26}$Al map \citep{Diehl1995} to study the Orion region in the context of our other observables. \citet{Diehl2002} and \citet{Diehl2004} studied the emission in Orion, and we build on their analysis in this paper. Additional statistical modeling of the $^{26}$Al emission's origins will be presented in Paper II. \\

\subsection{Other Observations}

We use the Planck $E(B-V)$ extinction map from \citet{Planck2011} to compare the 3D \citetalias{Leike2020} dust map with the 2D distribution of the dust. We use H$\alpha$ observations from \citet{Finkbeiner2003} to investigate Barnard's Loop and the surrounding excited gas. Figure \ref{fig:main_fig} shows a finding chart of the region with some of the observations superimposed: H$\alpha$, Planck $E(B-V)$, $^{26}$Al emission, and the stars featuring 3D velocity information from the OBP-B1 cluster. 

\begin{figure*}
    \centering
    \includegraphics[width=1.0\textwidth]{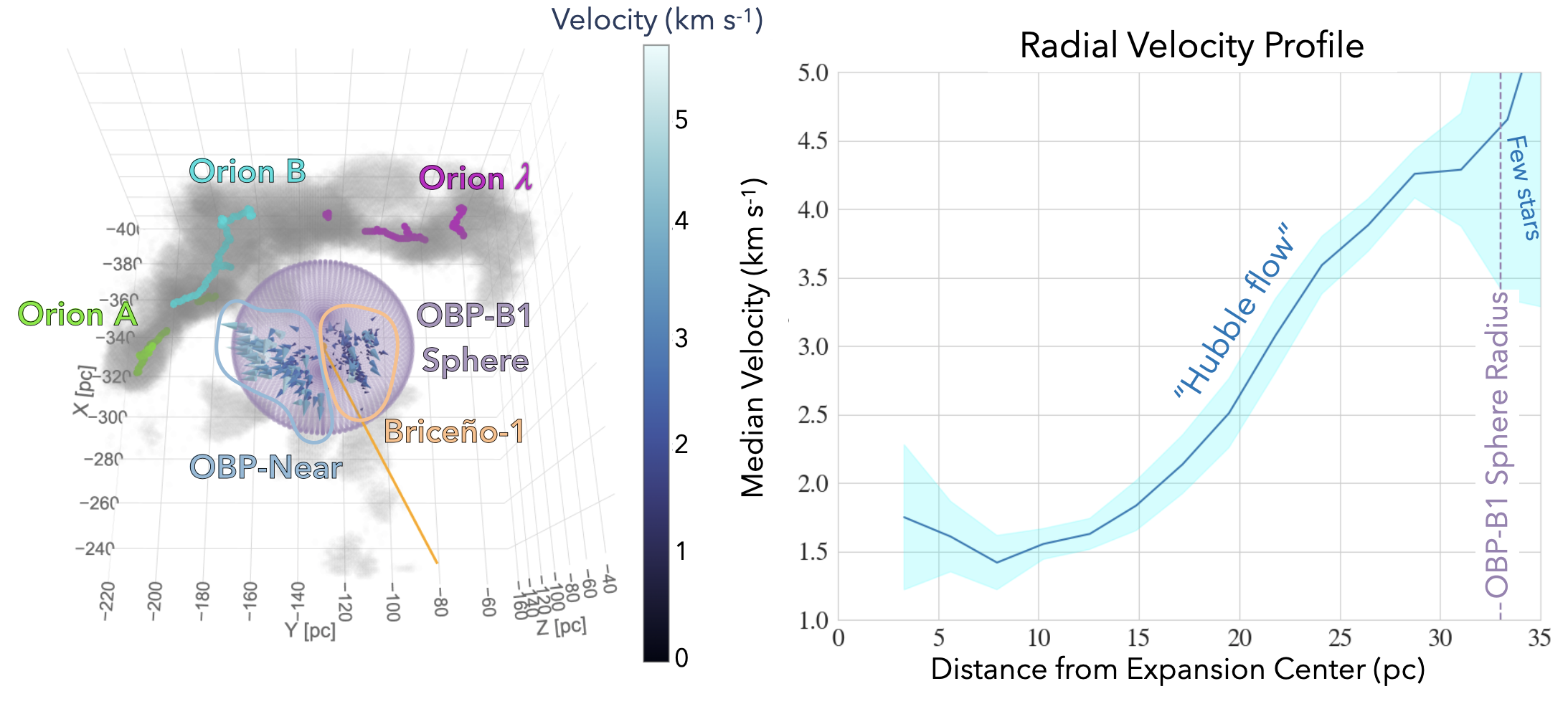}
    \caption{\textit{Left}: The 3D positions and velocities of stars alongside \citetalias{Leike2020}'s dust map in gray. Molecular cloud spines are shown in green, blue, and magenta for Orion A, B, and $\lambda$, respectively. Stars are colored by their expansion velocity relative to the expansion center of the cluster. The orange point at the end of the orange line indicates the expansion center. The orange line shows the direction to the sun. 
    \textit{Right}: Radial profile of velocities for all the stars in OBP-B1. The blue line corresponds to the median velocity at that distance from the expansion center, and the shaded blue region corresponds to 1$\sigma$ uncertainty (representing the dispersion in the 100 mock positions and 3D velocities calculated in Section \ref{subsec: cluster_expansion}. The dashed purple line indicates where the turnoff occurs, which matches the distance from the expansion center to the small foreground dust clouds. The interactive version of the left panel can be found \href{https://mfoley-astro.github.io/OrionPaper/orion_region_interactive.html}{here}
    \label{fig:stellar_motions}.} 
\end{figure*}

\section{3D Structures}
\label{sec: 3D_structures}

In our 3D map of the region (Fig. \ref{fig:3DFindingChart}), we combine the following data sets: the 3D dust map from \citetalias{Leike2020}; 3D position and velocity information for OBP-B1 from \citetalias{Swiggum2021} and \citetalias{Grosschedl2021}; and molecular cloud spines from \citetalias{Zucker2021}. Several features are immediately apparent, and the following will be referenced as the main structures throughout the paper (See Fig. \ref{fig:3DFindingChart} for the 3D structures):
\begin{enumerate}
  \item \textbf{Radial expansion of OBP-B1:} The members of OBP-B1 are moving coherently away from a single point that falls close to the center of Barnard's Loop on the sky. As shown in Figure \ref{fig:stellar_motions}, stars within $\sim 35$ pc of the OBP-B1 cluster's center exhibit a ``Hubble flow" profile: their velocity away from the projected cluster center increases linearly (from 1 km s$^{-1}$ up to 5 km s$^{-1}$) with distance from the center (see \citet{Kuhn2019} for more examples). The median expansion velocity is 2.1 km s$^{-1}$. See Section \ref{subsec: cluster_expansion} for more details.
  \item \textbf{Radial expansion of gas throughout the Orion cloud complex:} Beyond OBP-B1, YSOs studied in \citetalias{Grosschedl2021} serve as tracers of gas and also exhibit coherent radial motion away from the expansion center of OBP-B1. Some notable tracers in the expansion include IC2118, L1616, L1622, Orion B, and the head of Orion A. In this work, we transform the gas tracer velocities into the rest frame of OBP-B1. Gas velocities measured in this rest frame range from 2-10 km s$^{-1}$, with a median velocity of 5.7 km s$^{-1}$. \citetalias{Grosschedl2021} associate the expansion with the Orion X population, but this work argues feedback from both Orion X and OBP-B1 may have contributed to the expansion.
  \item \textbf{Sphere around OBP-B1 (OBP-B1 Sphere):} The OBP-B1 Sphere is a 33 pc sphere around the OBP-B1 cluster, denoting the extent of coherent stellar expansion. The sphere is centered at (x, y, z) = (-312, -130, -105) pc in Galactic Cartesian coordinates. In 3D, the OBP-B1 Sphere also intersects dust clumps in the foreground of OBP-B1 on the Sky, suggesting that feedback from OBP-B1 played a role in sweeping up the foreground dust. This sphere serves as a model for the sphere of influence of OBP-B1, even though the influence of the cluster is likely highly asymmetric due to the asymmetric distribution of dust around OBP-B1. 
  \item \textbf{Dust shell around OBP-B1 (Orion Shell):} The Orion Shell is a large, ellipsoidal shell of dust, roughly defined by the equations:
  \begin{align}
  \label{eq:orion_shell}
      x (pc) &= -317 + 40\cos{\theta}\sin{\phi}\\ \nonumber 
      y (pc) &= -120 + 65\cos{\theta}\sin{\phi}\\ \nonumber 
      z (pc) &= -115 + 50\cos{\theta}\sin{\phi}\\ 
      \nonumber
      \theta &\in [0, 2\pi], \phi \in [0, \pi]
  \end{align}
  
  Orion A, Orion B, Orion $\lambda$, foreground dust, and various regions of gas traced by \citetalias{Grosschedl2021}, such as L1616 and L1622, lie on the surface of this large cavity. The center of the ellipsoidal Orion Shell is offset from the expansion center of OBP-B1 by approximately 15 parsecs. In other words, the foreground dust falls $\sim 35$ pc away from the expansion center of OBP-B1, while the opposite side of the shell (where the molecular clouds lie) is $\sim 50-60$ pc away. 
\end{enumerate}

\section{Physical Conditions of Barnard's Loop}

\label{sec: Loop_conditions}
\citet{Reynolds1979} first found an expansion velocity for Barnard's Loop of $\sim 15$ km s$^{-1}$, which corresponds to a kinetic energy of at least $1.7 \times 10^{50}$ erg and an ionized gas mass of $8 \times 10^{4} \textrm{M}_{\odot}$. \citet{Reynolds1979} identified H$\alpha$ flux throughout the total Orion-Eridanus region of:

\begin{equation}
    F_{\alpha} = 2.5 \pm 0.7 \times 10^{-6} \textrm{ergs cm}^{-2} \textrm{s}^{-1}
\end{equation}

where the contribution from Barnard's Loop is $\sim 20 \%$, yielding 

\begin{equation}
    F_{\alpha, \textrm{Loop}} = 5 \pm 1.4 \times 10^{-7} \textrm{ergs cm}^{-2} \textrm{s}^{-1}.
\end{equation}

\citet{Heiles2000} identify a pressure of $P/k \approx 24,000 \textrm{cm}^{-3}$ K within the Loop. Additionally, they find that very small dust grains are more abundant in Barnard's Loop than in the HI throughout Orion. Densities in the Loop appear to range from 0.7 cm$^{-3}$ in the western arc to 3 cm$^{-3}$ closer to the central portion \citep{Madsen2006, ODell2011}. It is believed that the Loop is kept ionized by the most massive and luminous stars in Orion, such as the Belt stars \citep{Bally2008, ODell2011}. We note that the conditions within Barnard's Loop are still fairly uncertain. Indeed, even the previously quoted values are not consistent with each other. Densities of 0.7 cm$^{-3}$ to 3 cm$^{-3}$ at a temperature of $T \approx 6,000$ K yield pressures of $P/k \approx 4,000 - 18,000 \textrm{cm}^{-3}$ K, instead of the $P/k \approx 24,000 \textrm{cm}^{-3}$ K identified by \citet{Heiles2000}. 

Barnard's Loop has often been associated with the Orion-Eridanus superbubble, a large expanding structure in the vicinity of the Orion complex that subtends $\approx 20^{\circ} \times 40^{\circ}$ on the sky \citep{Reynolds1979}. The 3D geometry of the Orion-Eridanus superbubble remains unclear, but various studies indicate that it extends from a near distance of $\sim 150-200$ pc away to a far distance of $\sim 400-450$ pc \citep{Diehl2004, Ochsendorf2015, Joubaud2019}. This superbubble has a global expansion velocity of roughly 15 km s$^{-1}$ \citep{Reynolds1979, Bally2008}.

\citet{Reynolds1979} find that a series of supernovae, centered around $t \sim 2$ Myr ago and releasing a total energy of $\sim 3 \times 10^{52}$ ergs, could explain the expansion of the superbubble and, consequently, Barnard's Loop. It has been suggested that $\sim 10-20$ supernova explosions may have occurred in the Orion region over the past $\sim 12$ Myr. Consequently, a combination of nested shells may make up the superbubble \citep{Bally2008}. 

A work by \citet{Ochsendorf2015} looking at Planck and WISE data in vicinity of the Orion-Eridanus superbubble has presented an intriguing discovery: Barnard's Loop seems to be kinematically distinct from the Orion-Eridanus superbubble, implying the existence of a separate bubble associated with Barnard's Loop. Coupled with the shells discovered in \citet{Joubaud2019}, this discovery validates the picture presented in \citet{Bally2008} of nested shells throughout the Orion-Eridanus superbubble. The shells will eventually merge with the superbubble interior, replenishing the momentum and energy of the superbubble. However, a number of these shells, such as the Barnard's Loop bubble, still appear to be kinematically distinct. 

\citet{Ochsendorf2015} claim that the Barnard's Loop bubble is likely a SNR produced $\sim 3 \times 10^{5}$ yr ago and is associated with gas expanding at high velocity ( $\sim 100$ km s$^{-1}$). The shell of this bubble has a mass of $6.7 \times 10^{3}$ M$_{\odot}$, a radius of 35 pc \citep{Ochsendorf2015}, an average density of $\sim 1 \textrm{cm}^{-3}$, and a kinetic energy of $6.7 \times 10^{50}$ erg. The Barnard's Loop bubble is then embedded within the larger Orion-Eridanus superbubble and serves to replenish the superbubble's mass and energy. 

\citetalias{Grosschedl2021} and \citetalias{Kounkel2020} find that coherent velocities among gas and stars around Orion A and B also support the idea of a feedback-driven bubble, though both studies suggest that expansion began around 6 Myr ago, roughly an order-of-magnitude older than the age derived by \citet{Ochsendorf2015}. \citet{Tahani2022} observe that the 3D magnetic field morphology of Orion A is arc-shaped, indicating that such a bubble may play a large role in shaping the modern topology and magnetic environment of Orion A. Additionally, \citet{Grosschedl2020} observe a sharp ($\sim$ order-of-magnitude) increase in star formation rate around the Orion Nebula Cluster (ONC) in Orion A compared to the southern part of the cloud. The sharp increase in the star formation rate may be evidence of triggered star formation produced by an expanding Orion Shell. In the next section, we consider what the kinematics of stars and dense gas can tell us about the environment of Barnard's Loop. 

\section{3D Kinematics}
\label{sec: 3D_kinematics}
\subsection{Cluster Expansion}
\label{subsec: cluster_expansion}

To compute the radial velocity profile of OBP-B1, we treat each star as being drawn from a 6D Gaussian probability distribution reflecting their measurement uncertainties in both position (3D) and velocity (3D). We then randomly sampled from the 6D probability distribution 100 times to create 100 mock stellar populations. This Monte Carlo process permits us to see how robust certain observations are with respect to these uncertainties (while assuming those to be uncorrelated). The stars in each of the 100 mock populations are binned by their distance to the expansion center, and the median and $1 \sigma$ values for the velocities in each bin are shown in Figure \ref{fig:stellar_motions}.

The turn-off from a ``Hubble flow" appears to occur around 33 pc, primarily due to a lack of stars beyond this distance. Interestingly, this is also the distance from the OBP-B1 expansion center to the small foreground clouds shown in Figure \ref{fig:stellar_motions}, which were first identified by \citet{RezaeiKh2020}. The turn-off also matches the radius of the Barnard's Loop bubble (35 pc) identified in \citet{Ochsendorf2015}. Out to this distance, stellar expansion velocities grow linearly from 1 km s$^{-1}$ to 5 km s$^{-1}$, with a median expansion velocity of 2.1 km s$^{-1}$. 

The expansion is not perfectly clean, and a number of stars that do not move purely outward radially possess some faster velocities. Velocities that appear to ignore the radial expansion may be due to imperfect measurements of distance, proper motion, or radial velocity. Alternatively, they could represent stars migrating from another population that have been clustered with OBP-B1 due to the velocity- and position-based clustering algorithm. 

It is interesting to note that a remarkably similar velocity profile was observed in $\lambda$ Ori \citep[see Figure 2 in][]{ZamoraAviles_GravFeedback}, where the ``Hubble flow" profile extended to the edge of the $\lambda$ Ori ring. The $\lambda$ Ori ring is thought to be the result of a supernova that went off roughly 1 Myr ago \citep{Dolan2002}. However, as discussed in Section \ref{subsec: hii_region_expansion} and shown in \citet{Grudic2022}, such ``Hubble flow" profiles can be produced without supernovae through gas evacuation from early stellar feedback. 

\subsection{Gravitational Drag}
\label{subsec: gravitational_drag}

\begin{figure*}[ht!]
    \centering
    \includegraphics[width=1.0\textwidth]{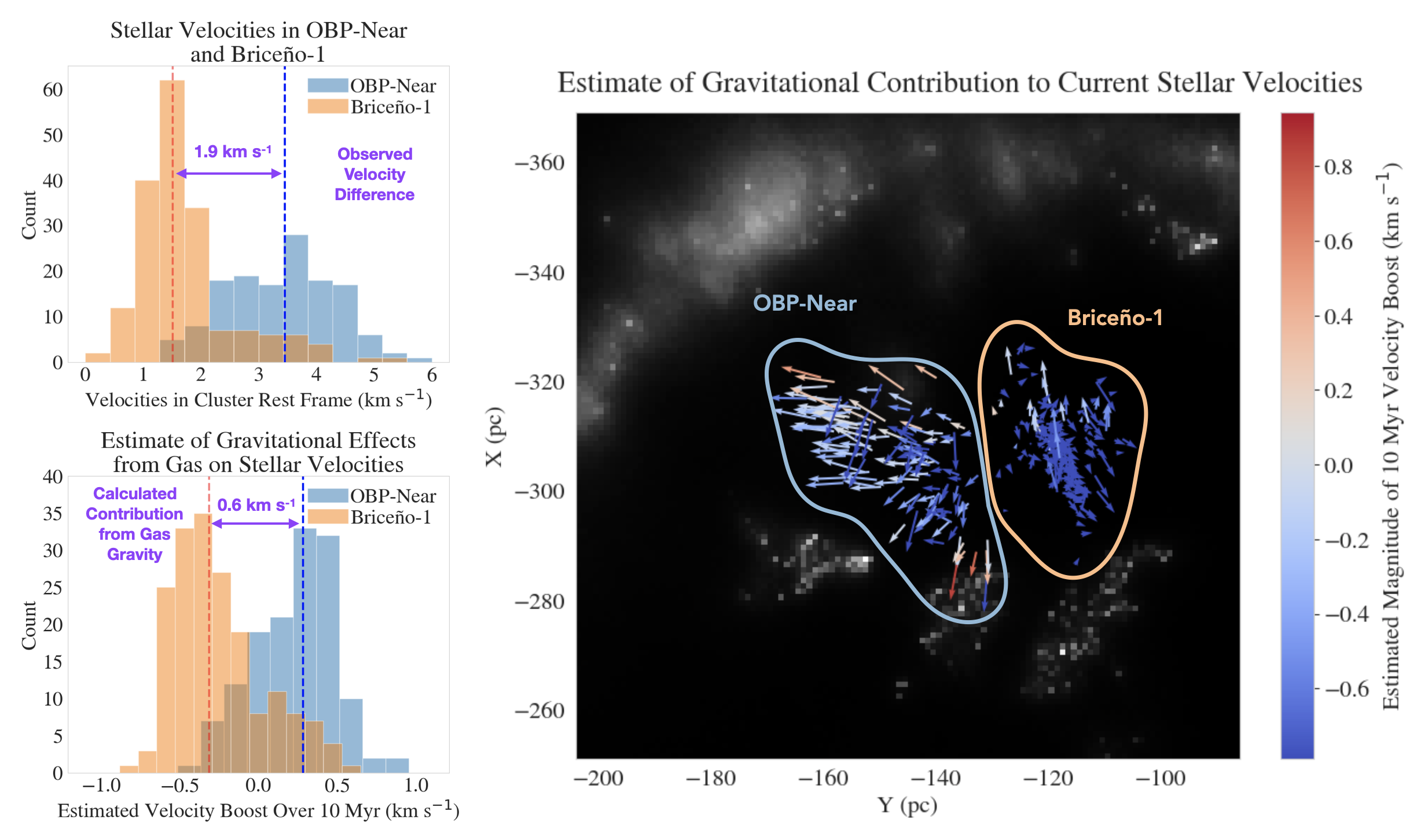}
    \caption{\textit{Top Left:} Histogram of stellar velocities in the rest frame of OBP-B1, separated into the stars from OBP-Near and Briceño-1. This is done since OBP-Near exhibits more coherent and faster motion than Briceño-1, so this serves as a natural division to test for gravitational drag. Taking the median velocities of both OBP-Near and Briceño-1 reveals a velocity difference of 1.86 km s$^{-1}$. \\
    \textit{Bottom Left:} An estimate of the velocity gained by gravitational acceleration from the nearby gas clouds, $\Delta v$ 
    over a period of $dt=10$ Myr, (see Eq.~\ref{eq: acc_v}). This calculation explains a velocity difference of 0.63 km s$^{-1}$, which is lower than the observed velocity difference by a factor of 3. \\
    \textit{Right:} Projection of dust density, stellar positions, and stellar velocity vectors on the XY Galactic Cartesian plane. The colors indicate the predicted difference in velocity due to the accumulated gravitational acceleration over 10 Myr by the dust and gas. Red indicates a star has likely been accelerated by the gas, and blue indicates a star has likely been decelerated by the gas.
    \label{fig:GravDrag}}
\end{figure*}

OBP-B1 exhibits an anisotropy in its expansion that does not appear to be present in the velocity structure of $\lambda$ Ori. \citet{ZamoraAviles_GravFeedback} suggest gas expulsion due to ionizing feedback from massive stars can ``flip" the gravitational potential; in other words, dense gas in an asymmetric expanding shell can drag stars outward by gravitational pull. While a perfectly symmetric shell would exert no gravitational force on stars within the shell, an asymmetric one can do so. An asymmetric gravitational force could then impart anisotropy in the velocities of stars in an expanding cluster. 

To see whether this ``gravitational drag" mechanism could have affected the kinematics of OBP-B1, we evaluate the gravitational effect of the surrounding gas and dust on the stars. This is done using the actual dust data from \citetalias{Leike2020} and not a model, allowing us to compute the true 3D gravitational acceleration vector for each star. We restrict ourselves to the gravitational acceleration since we do not have a reliable estimate for masses of individual stars in OBP-B1. As we have already converted the dust opacity in each voxel to a gas mass, we sum the gravitational acceleration from all of the voxels for each star as follows,

\begin{equation}
\label{eq: acc_vec}
    \textbf{a}_{i} = G \sum_{j} m_{j} \frac{\textbf{r}_{j} - \textbf{r}_{i}}{|\textbf{r}_{j} - \textbf{r}_{i}|^{3}}
\end{equation}
where $i$ corresponds to stellar indices and $j$ corresponds to voxels. For the purposes of this calculation, we ignore the stellar mass since the dust and gas mass strongly dominates. Computing the magnitude of the resulting acceleration vector gives the total gravitational acceleration on the star (see Section \ref{subsec: dust_calc} for the conversion of dust opacity into dust and gas mass). 

To determine the velocity boost/reductions accumulated from asymmetric gravitational acceleration, we first determine the angle $\theta$ between the gravitational acceleration vector on each star and that star's position vector measured from the expansion center of OBP-B1, as

\begin{equation}
\label{eq: acc_theta}
    \theta = \cos^{-1}(\frac{\vec{\mathbf{a}_{i}} \cdot \vec{\mathbf{r}_{i}}}{|\vec{\mathbf{a}}_{i}| |\vec{\mathbf{r}}_{i}|})
\end{equation}

Here we normalize each position vector, so $|\vec{\mathbf{r}}_{i}|\ = 1$ for all $i$. This calculation provides a measure of the alignment between the gravitational force from gas and dust on each star and the star's assumed expansion trajectory. 
Observations of the present situation cannot provide enough information to determine exactly how the dust and gas near OBP-B1 expanded over time. So, for simplicity, we consider the calculated ``present" acceleration vectors as a good first-order approximation of past acceleration, given that we assume that both the dust shell and the cluster are expanding. Consequently, we project the acceleration vector on to the position vector of the star to determine the component of acceleration contributing to radial expansion. This quantity is then multiplied by 10 Myr, the upper limit for the estimated age of OBP-B1, to evaluate how much such an acceleration would change the stellar velocities after 10 Myr:

\begin{equation}
\label{eq: acc_v}
    \Delta v = \cos(\theta){|\vec{\mathbf{a}}_{i}|}*(10 \,\textrm{Myr})
\end{equation}

If gravity is decelerating a star, this calculation will yield a negative number. Conversely, a positive result would indicate that gravity is accelerating a star. The results of this calculation are shown to the bottom left of Figure \ref{fig:GravDrag}. There is a clear distinction between the gravitational effects on OBP-Near, the population closer to dense gas and dust, and Briceño-1, the population farther away. We depict this in the image to the right of Figure \ref{fig:GravDrag}. ``Gravitational drag" appears capable of accounting for roughly 0.6 km s$^{-1}$ of velocity anistropy. This calculation only accounts for 1/3 of the observed 1.9 km s$^{-1}$ anisotropy, so other factors -- or a more sophisticated analysis -- are necessary to explain the full anisotropy.

\begin{figure*}
    \centering
    \includegraphics[width=0.75\textwidth]{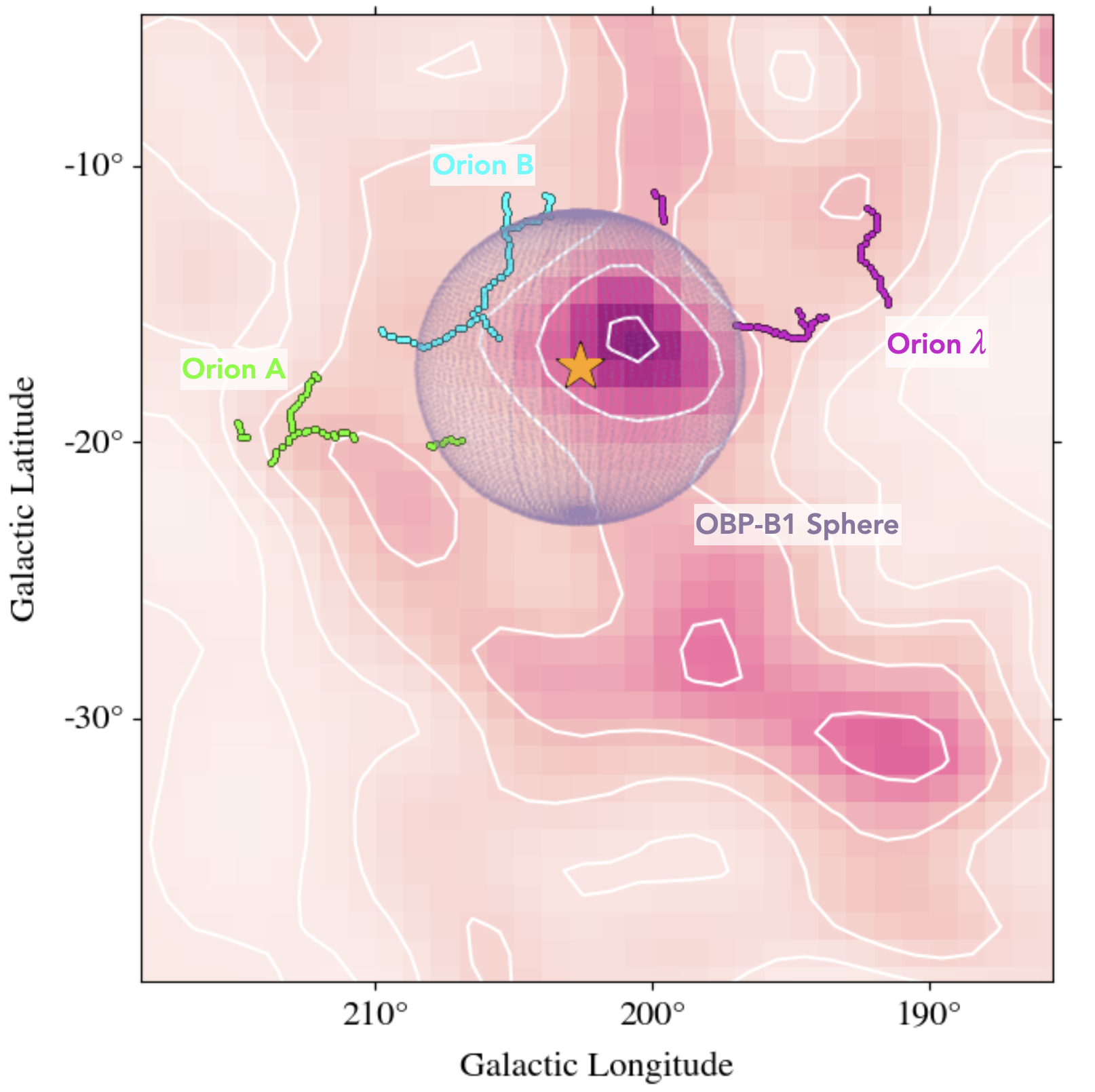}
    \caption{Contour map of $^{26}$Al emission in the Orion region. See \citet{Diehl2002} for further discussion of $^{26}$Al in Orion. The strongest enhancement of $^{26}$Al is found in very close proximity to the expansion center of OBP-B1. See Fig. \ref{fig:main_fig} for overlays of H$\alpha$ and dust extinction on top of the $^{26}$Al emission.
    \label{fig:Al26Contour}} 
\end{figure*}

The presence of the molecular clouds produces a tidal field that seems to explain at least some part of the velocity difference found between the stars in OBP-Near and Briceño-1. We have only performed a simple estimate, and more detailed physical analysis is outside the scope of this paper (\citet{Konietzka} will present a more sophisticated analysis), but modern simulations provide the ideal laboratory to determine if and how such a diverging stellar velocity distribution may have developed due to gravitational drag.

\subsection{What is the center of expanding gas in Orion?}
\label{subsec: yso_expansion}

Beyond OBP-B1, we also analyze the expansion of YSO clusters identified in \citetalias{Grosschedl2021}. As YSOs largely remain in their natal gaseous environment, YSO clusters serve as tracers for the gas dynamics in the Orion region. According to \citetalias{Grosschedl2021}, radial expansion throughout the Orion region can potentially be attributed to the recently discovered Orion X population (with additional contributions from the Orion OB1 association). Here, we instead evaluate the possibility that OBP-B1 serves as the true expansion center. 

In Figure \ref{fig:3DFindingChart}, {\it all} YSO clusters from \citetalias{Grosschedl2021} are placed into the rest frame of OBP-B1. The OBP-B1 expansion center also appears to function as a possible expansion center of the YSO gas tracers identified in \citetalias{Grosschedl2021}, in contrast to the Orion X expansion center hypothesized in that work. If the OBP-B1 expansion center also serves as the expansion center for large-scale dynamics in the region, it supports the claim that feedback from OBP-B1 played a large role in shaping the gas in the Orion region. However, a quantitative evaluation of possible expansion centers is still warranted. 

\citetalias{Grosschedl2021} report 3D velocity vectors for their gas tracers, which can be compared to position vectors of the YSO clusters measured from different centers of expansion. In the event of perfect radial expansion, the velocity vector and position vector of each YSO cluster should be perfectly aligned. Consequently, we can test the alignment between position and velocity vectors of the \citetalias{Grosschedl2021} gas tracers measured from both Orion X and the OBP-B1 expansion center. To do so, we calculated the average angular separation between the velocity vectors and position vectors to evaluate the possibility of radial expansion from different points.

\begin{equation}
\label{eq: YSO_pos}
    \textbf{r}_{i} = \textbf{x}_{i} - \textbf{x}_{c}
\end{equation}

Here $\textbf{x}_{c}$ is one of two expansion centers: OBP-B1 or Orion X. $\textbf{x}_{i}$ denotes the position of each YSO cluster in \citetalias{Grosschedl2021} in 3D space. Next, we simply calculate:

\begin{equation}
\label{eq: YSO_theta}
    \theta_{i} = \cos^{-1}(\frac{\vec{\mathbf{v}_{i}} \cdot \vec{\mathbf{r}_{i}}}{|\vec{\mathbf{v}}_{i}| |\vec{\mathbf{r}}_{i}|})
\end{equation}

where $\theta_{i}$ measures the separation between the position vector (measured from one of the two expansion centers) and the velocity vector for a single YSO cluster. Averaging the angles between velocity vectors and position vectors measured from each expansion center, we find an average angular separation of $25.42^{\circ}$ using OBP-B1 as the expansion center and $19.28^{\circ}$ using Orion X as the expansion center. In perfect radial expansion, the angular measure would be $0^{\circ}$. Therefore, both points --- the center of OBP-B1 and Orion X --- are reasonable candidates for the center of the large-scale expansion in Orion, though neither appears to be strongly preferred over the other. Given the uncertainties in the above estimates, we cannot definitively claim a common center for expansion. Rather, it seems likely that feedback from both Orion X and OBP-B1 contributed to the expansion of the clusters reported by \citetalias{Grosschedl2021}.

\section{Possible Supernovae}
\label{sec: possible_supernovae}
\subsection{How Many Supernovae Did OBP-B1 Produce?}
\label{subsec: possible_supernova_count}

\begin{figure*}[ht!]
    \centering
    \includegraphics[width=1.0\textwidth]{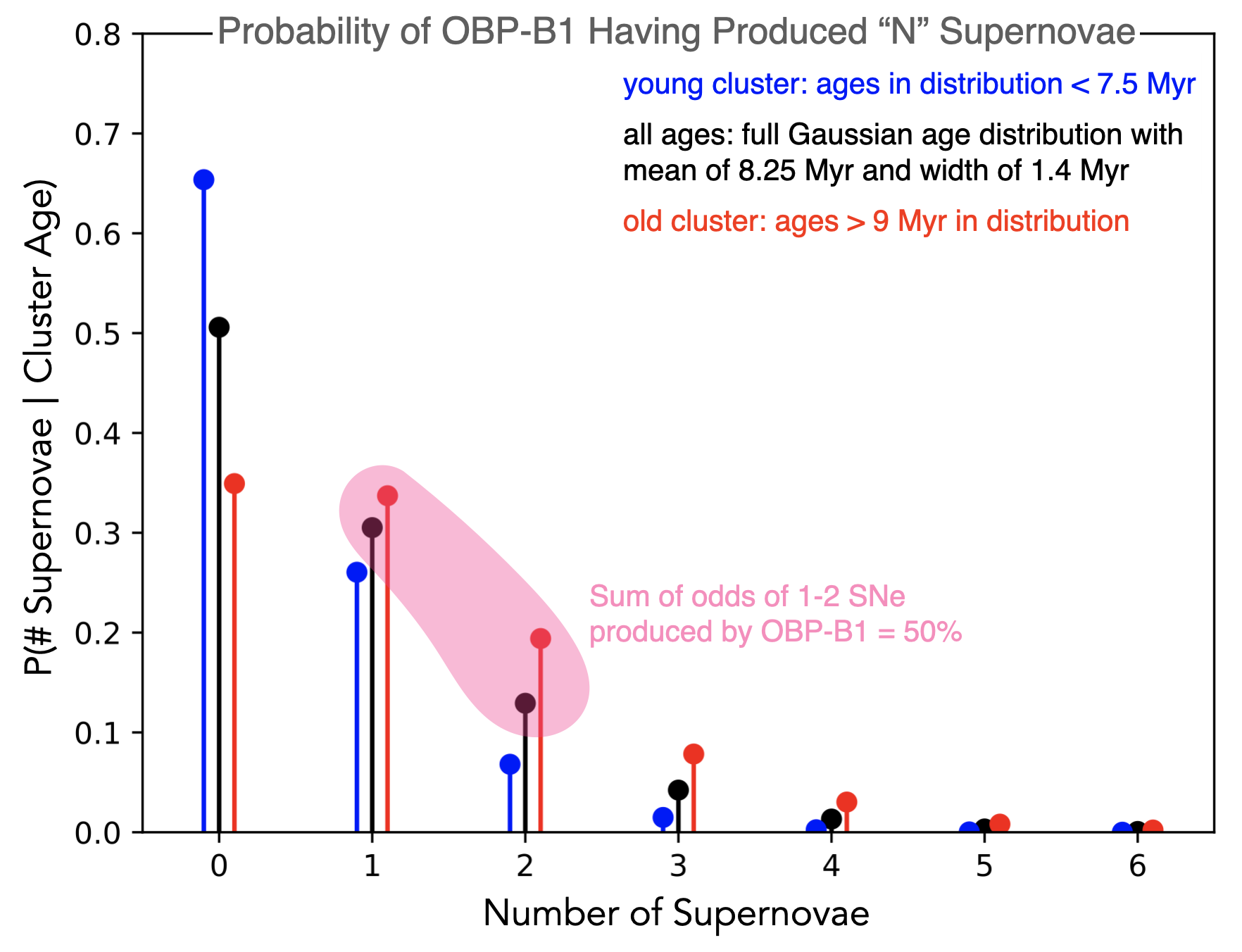}
    \caption{Probability of OBP-B1 producing a certain number of supernovae based on the mass and age of the cluster. Given the age uncertainties in the combined OBP-B1 cluster and the independent OBP-Near and Briceño-1 clusters, we consider the ``young" ($<$ 7.5 Myr old) and ``old" ($>$ 9 Myr old) scenarios separately. We argue that the cluster is more likely to fall on the older side, so the black and red lines are most relevant to the arguments of this paper. For the old scenario, there is a 65$\%$ chance of at least 1 supernova and 35$\%$ of no supernovae. For the full age range, the corresponding odds of OBP-B1 producing 1-2 supernovae are 45$\%$. 
    \label{fig:SNCount}} 
\end{figure*}

It is possible that OBP-B1 produced a supernova that strongly influenced the Barnard's Loop region. Figure \ref{fig:Al26Contour} shows a contour map of the $^{26}$Al emission in the region, indicating a strong enhancement very close to the expansion center of OBP-B1. $^{26}$Al has a half life of $\tau \sim 1$ Myr, so this emission traces particularly recent feedback events. This enhancement may well be due to other stars in the OB1a subgroup, so we investigate the possibility of a supernova from OBP-B1 with statistical modeling. We use Bayesian forward modeling introduced in \citet[][hereafter \citetalias{Forbes2021}]{Forbes2021} to calculate the probability of such an event using only the age and mass of OBP-B1. In Paper II, we use this model to further study the extended $^{26}$Al emission in the region. 

For this model, we use a mean age prior of $8.25 \pm 1.41$ Myr for OBP-B1. This is slightly older than the isochrone age of OBP-B1 (see Section \ref{subsec: cluster_formation} for more details), where fitting is done using PARSEC isochrones with EDR3 passbands \citep{Marigo2017}. However, 8.25 Myr is more consistent with the classical ages of the OB1a and OB1b subgroups to which OBP-Near and Briceño-1 belong \citep{Blaauw1964, Bally2008}. 

A Gaussian age spread is assumed for each star in the cluster, meaning that individual stellar ages are drawn from a Gaussian distribution centered around the mean age of 8.25 Myr. This width of this Gaussian distribution is a uniform prior from 0 to 2 Myr. In other words, the stars within our modeled cluster have an age spread of 2 Myr and an average age drawn from the Gaussian distribution of $8.25 \pm 1.41$ Myr. See \citet{Forbes2021} for more details. 

We then use $1 \times 10^3$ M$_{\odot}$ for the mass of our sampled cluster. This mass was calculated by producing a color-magnitude diagram for OBP-B1, matching each star to the closest point on the best-fit isochrone, and reading off the mass for each of the closest points on the isochrone. We then sum up the individual masses to arrive at a total mass for the cluster. Mass and age are the only two inputs for the model; the rest of the calculations rely on stellar evolutionary tracks and IMF sampling. It is important to note that the current membership list for OBP-B1, and even the classical Orion OB1 subgroups, are likely to be incomplete. These estimates may therefore be lower-bounds to true supernova rates from these clusters. 

Figure \ref{fig:SNCount} shows the results of predictions using the \citetalias{Forbes2021} methodology about the number of supernovae potentially produced by an OBP-B1-like cluster, depending on its age. The three possibilities are calculated by truncating the results based on the inputs: 1) using only results that had an initial age $> 9$ Myr old; 2) using all of the result, i.e. the full Gaussian prior; and 3) using only results for clusters had an initial age $< 7.25$ Myr old. As discussed previously, it is likely that OBP-B1 lies at the older end of the age range, so we particularly focus on the possibilities of an ``old" cluster ($> 9$ Myr old) or one with the full age prior. The model, as explained in Figure \ref{fig:SNCount}'s annotations, gives a probability of $\sim 50 \%$ that OBP-B1 produced 1-2 supernovae given these conditions (using either the full age prior or truncating the results to clusters $> 9$ Myr old). The full age distribution gives a probability of 45\% for 1-2 supernovae, while the old age truncation gives a probability of 65\% for 1-2 supernovae. Three or more supernovae appear to be quite unlikely for a cluster with the age and mass of OBP-B1. 

\begin{figure*}[ht!]
    \centering
    \includegraphics[width=1.0\textwidth]{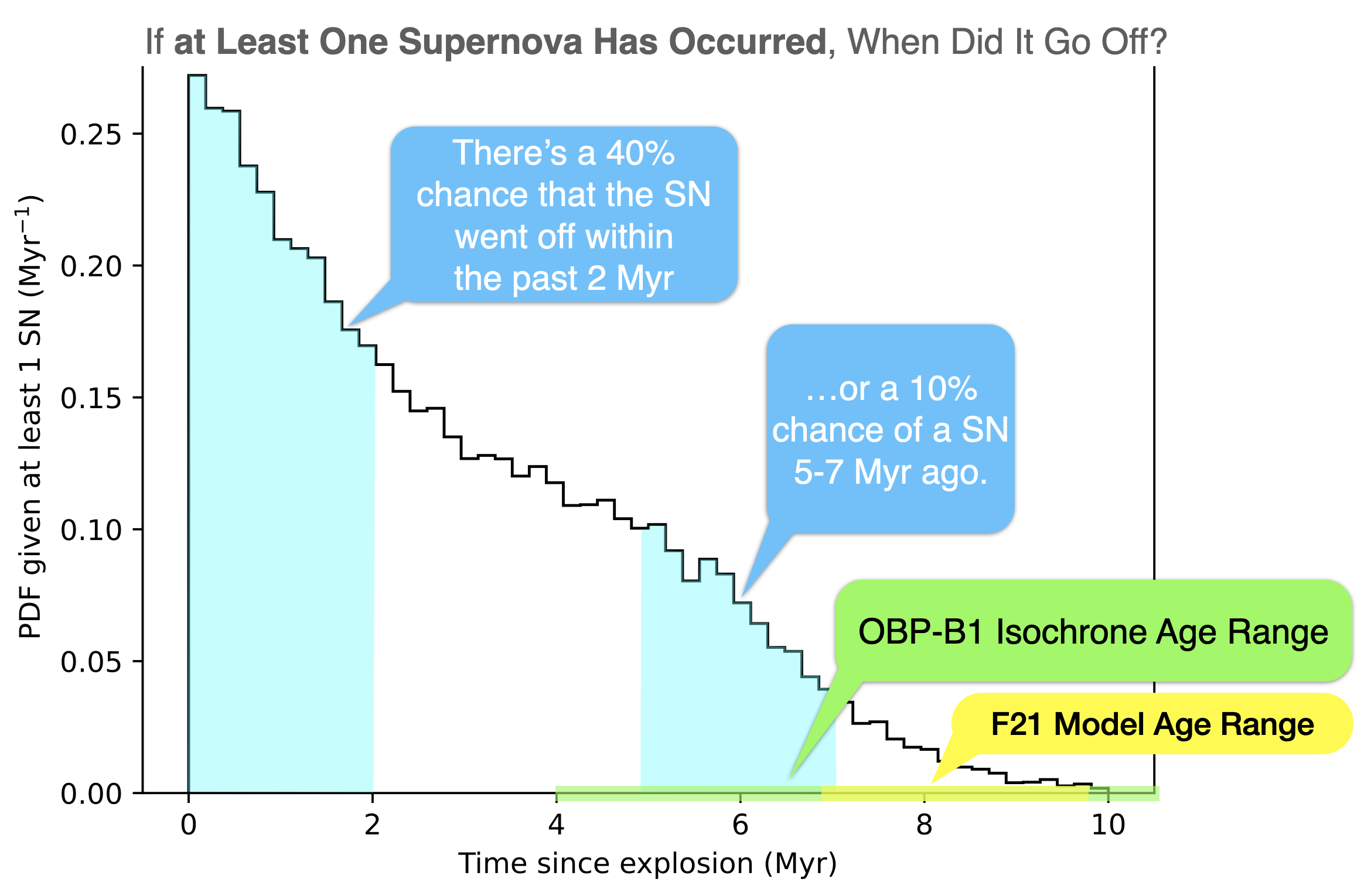}
    \caption{If we assume that OBP-B1 has produced at least one supernova, when did that supernova (or supernovae) go off? This figure presents the PDF of time elapsed since a supernova explosion based on the Bayesian modeling by F21. More recent supernova explosions are heavily favored by the model. For example, there is roughly a $40\%$ chance of a supernova exploding 0-2 Myr ago in OBP-B1, which is 4 times more likely than a supernova exploding 5-7 Myr ago. 
    \label{fig:SNTiming}} 
\end{figure*}

\subsection{When Did Supernovae from OBP-B1 Go Off?}
\label{subsec: possible_supernova_timing}
If we \textit{assume} that at least one supernova goes off in OBP-B1, the \citetalias{Forbes2021} model also predicts its timing. Figure \ref{fig:SNTiming} shows the probability density function of supernova explosion times, given that at least one explosion occurs. The model predicts a much higher likelihood for recent supernova explosions (past few Myr) than for older ones (5-7 Myr as suggested by \citetalias{Grosschedl2021} and \citetalias{Kounkel2020}). These results support the timing proposed by \citet{Ochsendorf2015}, who estimate that a supernova producing Barnard's Loop bubble would have needed to explode around 0.3 Myr ago. As a result, we find that either the 6 Myr age estimates for supernova timing suggested by \citetalias{Grosschedl2021} and \citetalias{Kounkel2020} are overestimates, or, more probably, those large-scale analysis are also tracing the effects of older feedback events in the Orion complex that did not originate in OBP-B1.  

\subsection{Could Supernovae from OBP-B1 have produced the Orion Shell?}
\label{subsec: obp_b1_momentum}
Let us consider the momentum needed to create the Orion Shell observed around OBP-B1. We estimate the total gas mass of the Orion Shell, based on the 3d dust observations, as follows:
\begin{equation}
    M = \sum_{i=1} dm_i = \mu \sum_{i=1} n_i dV
\end{equation}

where $\mu$ is mean molecular weight, $n_{i}$ is the number density per voxel, and $dV = 1 \rm{pc}^{3}$ is the volume of each voxel. Fixing $\mu = 1.37 \times m_{p}$,
we find that the Orion Shell has a mass of M = $1.4 \times 10^5$ M$_{\odot}$. However, we note that the Orion Shell includes Orion B, Orion Lam, and most of Orion A, clouds that lie on the edge of the \citetalias{Leike2020} map. Our calculation of $1.4 \times 10^5$ M$_{\odot}$ is a rudimentary mass estimate of the entire shell, but it may in fact be a significant underestimate due to a different dust-to-gas ratio or additional gas not traced in the dust map. 

Each supernova involved in producing a superbubble can be expected to inject, on average, roughly $3 \times 10^5$ M$_{\odot}$ km s$^{-1}$ of momentum \citep{El-Badry2019}. The expansion velocity of the Orion Shell remains uncertain, but tracers from \citetalias{Grosschedl2021} lying on the edge of the shell are moving radially outwards (in the reference from of OBP-B1) with velocities of $3-8$ km s$^{-1}$. Tracers closest to the molecular clouds are moving with velocities of $5-7$ km s$^{-1}$, while L1622 is expanding into low-density material at velocity of 8.4 $5-7$ km s$^{-1}$. 

These velocities are all relatively close to the turbulent velocity dispersion of the ISM. The Milky Way is known to have a velocity dispersion of $\sim 7$ km/s in the cold neutral medium \citep{Heiles2003}. Extragalactic measurements find a very similar result. The PHANGS-ALMA survey analyzed CO (2--1) data from 80 galaxies, finding that the average mass-weighted velocity dispersion in the disks at 150 parsec scales is $5.8^{+1.9}_{-1.5}$ km s$^{-1}$ \citep{Sun2022}. With an expansion velocity close to the turbulent velocity dispersion of the ISM, the denser parts of the Orion Shell appear close to dissolving into the surrounding ISM.  

Consequently, we use 7 km s$^{-1}$ as a reasonable value for the expansion velocity of the Orion Shell and the turbulent velocity dispersion of the cold neutral medium. Multiplying the mass in the Orion Shell by this velocity yields a total momentum of $p \sim 8.4 \times 10^{5}$ M$_{\odot}$ km s$^{-1}$. Ignoring the momentum contributions from stellar winds and other types of stellar feedback, 2-3 supernovae can account for the momentum. However, stellar winds and other channels of stellar feedback have been found to inject momentum comparable to a supernova over their full lifetime \citep{Grudic2022}. Thus, the addition of perhaps only 1-2 supernovae to other forms of stellar feedback from OBP-B1 could inject enough momentum to produce the Orion shell.  

We can also remain agnostic about the current expansion velocity of the shell and instead perform an alternative momentum estimate using only the radius of the shell ($R_{shell}$), the mass of the shell ($M$), and the expansion time of the gas ($t_\mathit{expansion}$):

\begin{equation}
    p = \eta \frac{MR_{shell}}{t_\mathit{expansion}}
\end{equation}

Here $\eta$ is a scaling factor accounting for the radius growth of an expanding supernova remnant, $R \propto t^{\eta}$, at late stages. Values for $\eta$ have been determined by theoretical models and simulations of supernova remnants \citep[see e.g.][]{Kim2015, El-Badry2019}. For a shell powered by multiple supernovae (as is likely the case for the Orion Shell), \citet{El-Badry2019} find $eta \approx 0.5$ \citep{Bialy2021}.

With values of $R_{shell} = 60$ pc (the radius at which most of the mass in the shell resides), $M = 1.4 \times 10^5$ M$_{\odot}$, $\eta = 0.5$, and and $t_\mathit{expansion} = 6$ Myr (the time at which large-scale gas expansion seems to begin), we obtain $p \sim 7 \times 10^{5}$ M$_{\odot}$ km s$^{-1}$, in good agreement with out first estimate of momentum. 

Stellar winds, radiation pressure, and protostellar outflows all may have contributed early momentum to the gas surrounding OBP-B1, reducing the contribution from supernovae necessary to reach the current state. Additionally, much momentum will go into low-density gas that is quickly accelerated away from the site of explosion. Momentum injected into the dense shell only accounts for a fraction of the total momentum injected by a supernovae, so more supernovae will be required than either of the previous two calculations imply. Nevertheless, OBP-B1 appears capable of supplying the right order of magnitude of momentum to explain the expansion of the Orion Shell. 

Thus far in this section, we have only considered the possibility that OBP-B1 does produce at least 1 supernova, though the model suggests that there is a $50 \%$ chance for the full prior model that OBP-B1 produces no supernovae at all. If a supernova does not occur, the dynamics would be dominated by other forms of stellar feedback, and nearby supernovae from other clusters would likely have helped to form the Orion Shell.

\section{Formation Timeline}
\label{sec: timeline}
\subsection{OBP-B1 Formation}
\label{subsec: cluster_formation}

\begin{figure*}
    \centering
    \includegraphics[width=1.0\textwidth]{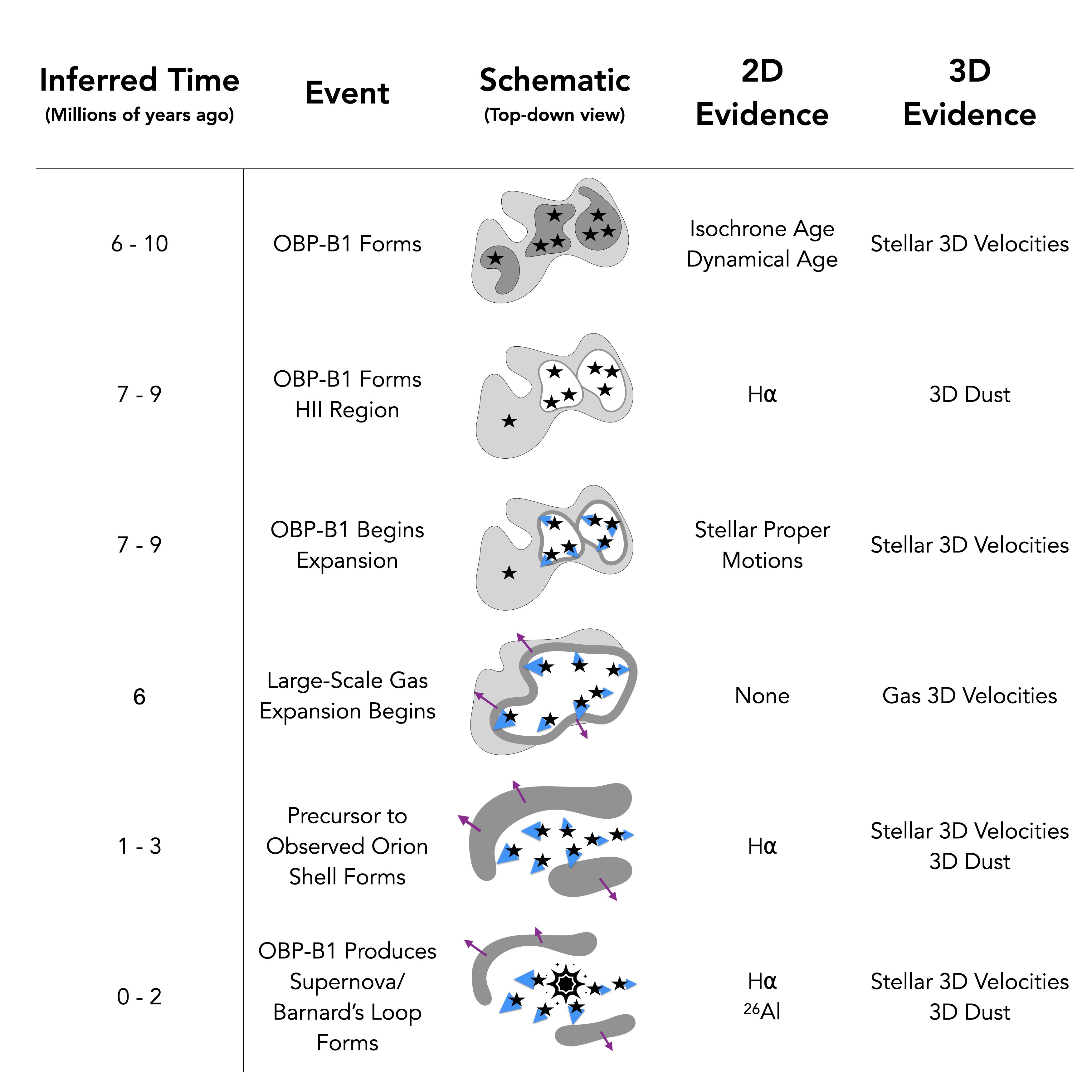}
    \caption{Timeline of major events in the Orion region. The left column indicates the rough timing for each event. The middle column shows a schematic of the Orion region over time (viewed from the Galactic plane looking down towards Orion). The right most columns indicate the 2D and 3D datasets that support the timing and schematic of each event. "Gas 3D Velocities" refers to data from \citetalias{Grosschedl2021}, "Stellar 3D Velocities" refer to our measurements of OBP-B1 and those of \citetalias{Swiggum2021}, and "3D dust" refers to the map from \citetalias{Leike2020}.
    \label{fig:orion_timeline}} 
\end{figure*} 

Here we discuss a likely timeline for the formation of Barnard's Loop bubble (see Fig. \ref{fig:orion_timeline}). The first event in our timeline for Barnard's Loop bubble is the formation of OBP-B1. The OBP-B1 cluster is composed of two groups (OBP-Near and Briceño-1) identified via velocity clustering, and fitting isochrones to each group independently reveals that OBP-Near may have a slightly younger age than Briceño-1. In \citetalias{Swiggum2021}, considering OBP-Near and Briceño-1 separately yields ages of $t_{iso} = 6.2^{+3.5}_{-2.4}$ Myr and $t_{iso} = 9.0^{+4.0}_{-3.4}$ Myr, respectively. Fitting an isochrone to the combined OBP-B1 population reveals an age of $t_{iso} = 6.8^{+3.6}_{-2.9}$Myr. The combined group OBP-B1 is considered throughout the paper, but the possibility of independent formation events is discussed when applicable.

By associating each star with its nearest point on the combined isochrone fit, we obtain a stellar mass of $\sim $1000 M$_{\odot}$ for OBP-B1. Since the cluster is radially expanding outward, extrapolating backwards in time using the present-day velocities indicates that the cluster likely began expanding 7.7-9.6 Myr ago. The dynamical age is slightly older than the combined isochrone age of $6.8^{+3.6}_{-2.9}$Myr. However, a cluster clearly cannot begin expanding before it is formed. Consequently, we suggest that the age of OBP-B1 falls on the older end of the isochrone age. However, we discuss alternative hypotheses and considerations regarding the possibility of two separate formation events for OBP-Near and Briceño-1 in Section \ref{subsec: alternatives}. 

Much of our timeline makes use of close comparison to simulations. In particular, we reference new STARFORGE simulations that resolve the formation of individual stars \citep{Grudic2021_STARFORGE, Grudic2022}. Studying a giant molecular cloud with initial mass $2\times10^{4}$ M$_{\odot}$ and radius of 10 pc, \citet{Grudic2022} find a final accreted stellar mass of $1600$ M$_{\odot}$ after 9 Myr of evolution. Given the close agreement in mass and age of the simulated cluster with OBP-B1, we discuss the simulated cluster as a possible analog for OBP-B1. 

The \citet{Grudic2022} simulation is a full radiation-magnetohydrodynamic simulation run using the GIZMO code with prescriptions for radiative feedback, protostellar jets, stellar winds, and supernovae. The simulation is run using $\sim$ 25 million $10^{-3}$ M$_{\odot}$ gas cells. See \citet{Grudic2021_STARFORGE} and \citet{Grudic2022} for more details on the physics prescriptions implemented in the simulation. 

\subsection{HII Region Formation and Expansion of OBP-B1}
\label{subsec: hii_region_expansion}
As shown in Figure \ref{fig:stellar_motions}, OBP-B1 exhibits a coherent radial expansion centered on a single point. The center is located close to the middle of a large dust shell, the Orion Shell, on which lie the three Orion molecular clouds. The Orion Shell also features the foreground dust recently discovered by \citet{RezaeiKh2020}. The OBP-B1 Sphere serves as an approximation of the sphere of influence of OBP-B1, which we determine to be $\sim 33$ pc in radius from the stellar kinematics. However, the effects of OBP-B1 on the surrounding gas likely happened in an asymmetric fashion, and the OBP-B1 Sphere should not be taken as a literal boundary for feedback from OBP-B1. No dust is observed on the right side of the figure, indicating a blowout from the feedback event that caused the cluster expansion. It is interesting to note that the cluster exhibits a spherical expansion similar to a ``Hubble flow" (see \citet{Kuhn2019} for other examples) in which stars farthest from the point of expansion travel the fastest. See Section \ref{subsec: cluster_expansion} for further details. In light of this velocity profile that extends to the boundaries of the large Orion Shell dust cavity (Figure \ref{fig:3DFindingChart}), we suggest that OBP-B1 produced feedback processes that began to remove gas from the vicinity of the cluster a few Myr after cluster formation. Eventually, there was not enough gas mass to keep OBP-B1 gravitationally bound and the cluster began to expand radially outward. Such a mechanism was observed in the simulation from the STARFORGE collaboration \citep{Grudic2022, Guszejnov2021}. 

In the \citet{Grudic2022} simulation, the star cluster begins forming at $t = 1$ Myr, has accreted half of the final stellar mass by $t = 5$ Myr, and experiences its first supernova at $t = 8.3$ Myr. However, at the time of the first supernova, most of the gas around the cluster has already been evacuated due to other stellar feedback processes, forming a cavity of gas roughly 20 pc in radius. Early stellar feedback processes include stellar winds, radiation pressure, UV-ionization, and protostellar jets. After roughly half of the stellar mass has been accreted by $t = 5$ Myr, feedback quickly reduces the bound gas mass within a few Myr. The dispersal of gas serves to unbind the central cluster and produce a ``Hubble flow" velocity profile by $t = 8$ Myr (see Figure 4 of \citet{Grudic2022}) that is remarkably similar to the one observed in OBP-B1 (Figure \ref{fig:stellar_motions}). The expanding gas fronts outrun the expanding cluster, creating a dense shell surrounding the stars. Such a shell is precisely what is observed around OBP-B1 (see Fig. \ref{fig:3DFindingChart}).

The expulsion of gas and expansion of the cluster occurs before the onset of the first supernova in the simulation, suggesting that supernovae are not required in order to produce the kinematics observed in OBP-B1. The ionized gas mass in the simulation also grows quickly between $t = 4$ Myr and $t = 8$ Myr after the first stars of $> 20$ M$_{\odot}$ form and begin to contribute significant ionizing luminosity. At the end of the simulation, ionized gas accounts for 20\% ($\approx$ few thousand M$_{\odot}$) of the total cloud mass. This ionized gas mass is roughly a factor of 20--30 lower than the $8 \times 10^{4} M_{\odot}$ of ionized gas in Barnard's Loop. As a result, we propose that OBP-B1 formed an HII region and expelled enough bound gas to unbind itself within a few Myr of formation, around 7--9 Myr ago, though this HII region has not been the dominant source of ionization in the Orion complex.  

\subsection{Large-Scale Gas Expansion}
\label{subsec: gas_expansion}

\citetalias{Grosschedl2021} and \citetalias{Kounkel2020} find that large-scale expansion in the Orion complex began around 6 Myr ago. \citetalias{Grosschedl2021} traces the gas and \citetalias{Kounkel2020} traces the stars, but both studies linearly extrapolate current velocities back in time to arrive at 6 Myr for the beginning of large-scale expansion. While these velocities likely did not respond constant throughout expansion, implying that a linear extrapolation is insufficient and 6 Myr may not be the true beginning of expansion, the independent measurement of 6 Myr from two studies using different tracers is highly suggestive. According to our modeling (see Section \ref{sec: possible_supernovae}), it is unlikely that OBP-B1 produced any supernovae that could trigger this expansion 6 Myr ago. The cluster is also not massive enough to influence the dynamics of the entire Orion region through winds, radiation pressure, or other forms of stellar feedback alone. 

Nevertheless, our analysis in Section \ref{subsec: yso_expansion} finds that the large-scale gas expansion can be traced to an origin close to both OBP-B1 and Orion X. If OBP-B1 did not produce the feedback necessary to begin large-scale expansion 6 Myr ago, it is possible that Orion X or another nearby cluster did produce a major feedback event at that time. Such an event would serve to assist in the unbinding of OBP-B1 and expedite the formation of a massive, dense shell, such as the precursor to the current Orion shell, a few Myr later. As this possibility ties together the work of \citetalias{Grosschedl2021}, \citetalias{Kounkel2020}, and this paper, we argue that large-scale expansion did begin 6 Myr, driven by supernovae outside of, but close to, OBP-B1.

\subsection{Shell Formation}
\label{subsec: shell_formation}

In the observations, foreground dust falls roughly 35 pc away from the expansion center of OBP-B1. Orion A, B and $\lambda$ reside closer to 50-60 pc away from the expansion center.  In the simulation, the gas cavity produced by feedback only reaches a radius of roughly 20 pc (measured by the half-mass radius of the cloud) by the time of the first supernova , $\sim 7$ Myr after the cluster began forming. At this time in the simulation just prior to the first supernova, the gas possesses a velocity dispersion of $\sigma \sim 10$ km s$^{-1}$. Assuming this gas experiences some deceleration as it expands into the ambient ISM, the gas cavity present at the end of the simulation could reach the size of the observed Orion Shell around OBP-B1 a few Myr later, even without a supernova. In the simulation, a supernova at $t = 8.3$ Myr ($\sim 7$ Myr after the cluster began forming) quickly increases the gas velocity dispersion and the size of the gas cavity, bringing the cavity into closer agreement with the observed Orion Shell. While it is not possible to definitively say if a supernova occurred in OBP-B1 through this comparison of simulation and observation, the timescales suggest that a supernova is a reasonable cause for the observed size of the Orion Shell around OBP-B1.

Consequently, we suggest that a similar physical mechanism occurred with OBP-B1 (see Figure \ref{fig:orion_timeline} for a complete breakdown). A few Myr after the first stars in OBP-B1 were born, stellar feedback expelled most of the gas from OBP-B1's natal environment. The energy from stellar feedback increased the gravitationally unbound gas mass and decreased the bound gas mass, which led to the formation and expansion of a dust shell and rapid unbinding of the cluster. Large-scale feedback in the region likely accelerated gas expansion around 6 Myr ago. After a few Myr, a shell around OBP-B1 with a radius of $\sim$ 10-20 pc had formed. At this point, it was large enough to encompass some of the other massive stars in Orion OB1 association. This shell then continued expanding into the ambient ISM, powered by winds, radiation pressure, and photoionization from growing portions of the Orion OB1 association. 

\subsection{Supernova(e)}
\label{subsec: supernova_timeline}

As discussed in Section \ref{sec: possible_supernovae}, our modeling finds that OBP-B1 is capable of producing 1-2 supernovae that likely exploded in the past few Myr. These rates are consistent with the amount of momentum needed to produce the Orion shell. Furthermore, our modeling supports the picture proposed in \citet{Ochsendorf2015}, in which a supernova went off $\sim 3 \times 10^{5}$ years ago in the vicinity of Barnard's Loop and produced a Barnard's Loop bubble kinematically distinct from Orion-Eridanus.  

Consequently, we argue that OBP-B1 produced at least one supernovae $\sim 3 \times 10^{5}$ years ago that accelerated the expansion of the large-scale gas shell in the region, leading to the current structures of the Orion Shell and Barnard's Loop. If OBP-B1 did indeed form 7-10 Myr ago, this timing is also consistent with the STARFORGE simulation, where the first supernova is produced 7 Myr after the cluster began forming.

The current Orion Shell features all three major molecular clouds in Orion: Orion A, Orion B, and Orion $\lambda$. This timeline predicts that the clouds on the Orion Shell should be in a period of strongly triggered star formation as a result of the swept-up gas from a supernova in OBP-B1. This is consistent with the enhancement of the SFR found in Orion A by \citet{Grosschedl2020}. Further work is necessary to explore this prediction throughout the Orion complex. 

\subsection{Alternative Hypotheses} 
\label{subsec: alternatives}

The dynamical and isochrone ages of OBP-B1 present a few difficulties to the proposed timeline. For example, OBP-Near and Briceño-1 appear to be clusters in slightly different evolutionary stages \citep{Swiggum2021} and members of different subgroups in the Orion OB1 association, suggesting that OBP-B1 ought not to be treated as a single cluster. The dynamical age of OBP-B1 is also older than the isochrone age, a clear impossibility for our proposed timeline where OBP-B1 becomes unbound after formation. Consequently, we consider a few different alternative hypotheses:
\begin{enumerate}
    \item OBP-B1 began expanding very shortly after formation. OBP-B1 likely would not expel enough gas to unbind itself at that time, so the source of feedback discussed in Section \ref{subsec: gas_expansion} may be fully responsible for unbinding OBP-B1. This would raise the possibility that OBP-Near and Briceño-1 could have formed separately in a large, shared gas reservoir, but the dispersal of this reservoir served to unbind both clusters and generate expansion from a consistent center.
    \item OBP-Near and Briceño-1 did not begin expanding from a single point, but rather possess two different ages and have inherited motions from separate events. One possibility is that there are two expansion centers: one at the center of OBP-Near and one at the center of Briceño-1. Briceño-1 may have formed first, expelled gas via stellar feedback, and began its own dispersal. The expanding gas consequently could have formed OBP-Near. 
    \item The stars have decelerated over time. The deceleration could be due to N-body interactions within OBP-B1 or interactions with other nearby star clusters. This hypothesis seems unlikely, since both possibilities would result in less coherent motion than is currently observed in OBP-B1.
\end{enumerate}

In support of hypothesis 1, \citetalias{Swiggum2021} found differences in the Class II disk fraction of stars in OBP-Near and Briceño-1, suggesting that OBP-Near is indeed at a younger evolutionary state. The sequential star formation picture in Orion proposed by \citet{Blaauw1964} also suggests that OBP-Near (part of the OB1b subgroup) and Briceño-1 (part of the OB1a subgroup) did form at different times. Finally, OBP-Near possesses higher velocities (relative to the expansion center of the combined OBP-B1 group), supporting the idea that OBP-Near may have been formed from gas expanding away from Briceño-1.

If OBP-Near and Briceño-1 are indeed two different clusters, we argue that our timeline should not be altered significantly. Our modeling and estimates of feedback effects should not depend on the distribution of stars into any number of clusters. Rather, the important quantities are the age and mass of the total stellar content in a region. Distinct ages for OBP-Near and Briceño-1 would likely change the timing and count of supernovae predicted in Section \ref{sec: possible_supernovae}, but these changes should be minor. This is because Briceño-1 would possess an older age and OBP-Near would possess a younger age than the value of $8.25 \pm 1.41$ Myr assumed in the model for the combined OBP-B1 cluster. As a consequence, Briceño-1 would be more likely to produce a recent supernova, even though OBP-Near would be less likely to produce one.  

We argue hypothesis 2 is even less plausible than hypothesis 1. It seems unlikely that a cluster of stars with the coherent motions of OBP-Near would form so uniformly in an expanding gas shell, especially if it is produced by only a few hundred solar masses of stars in Briceño-1. Additionally, the coherent radial expansion of OBP-B1 is difficult to explain if the formation events of OBP-Near and Briceño-1 are unrelated. 

Instead, we argue that OBP-B1 should be treated as a single cluster that was responsible for its own dispersal shortly after formation. Alternatively, strong external feedback may have contributed to gas expulsion that unbound OBP-B1. In the case of external feedback, distinct formation events for OBP-Near and Briceño-1 remain a possibility, though it remains less clear how the formation of the coherent shell of dust centered on OBP-B1 would proceed. Further work and comparisons with simulations will help to disentangle the various possibilities.

\section{Connecting the Timeline to Barnard's Loop}
\label{sec: barnard_loop_formation}

\begin{figure*}
    \centering
    \includegraphics[width=1.0\textwidth]{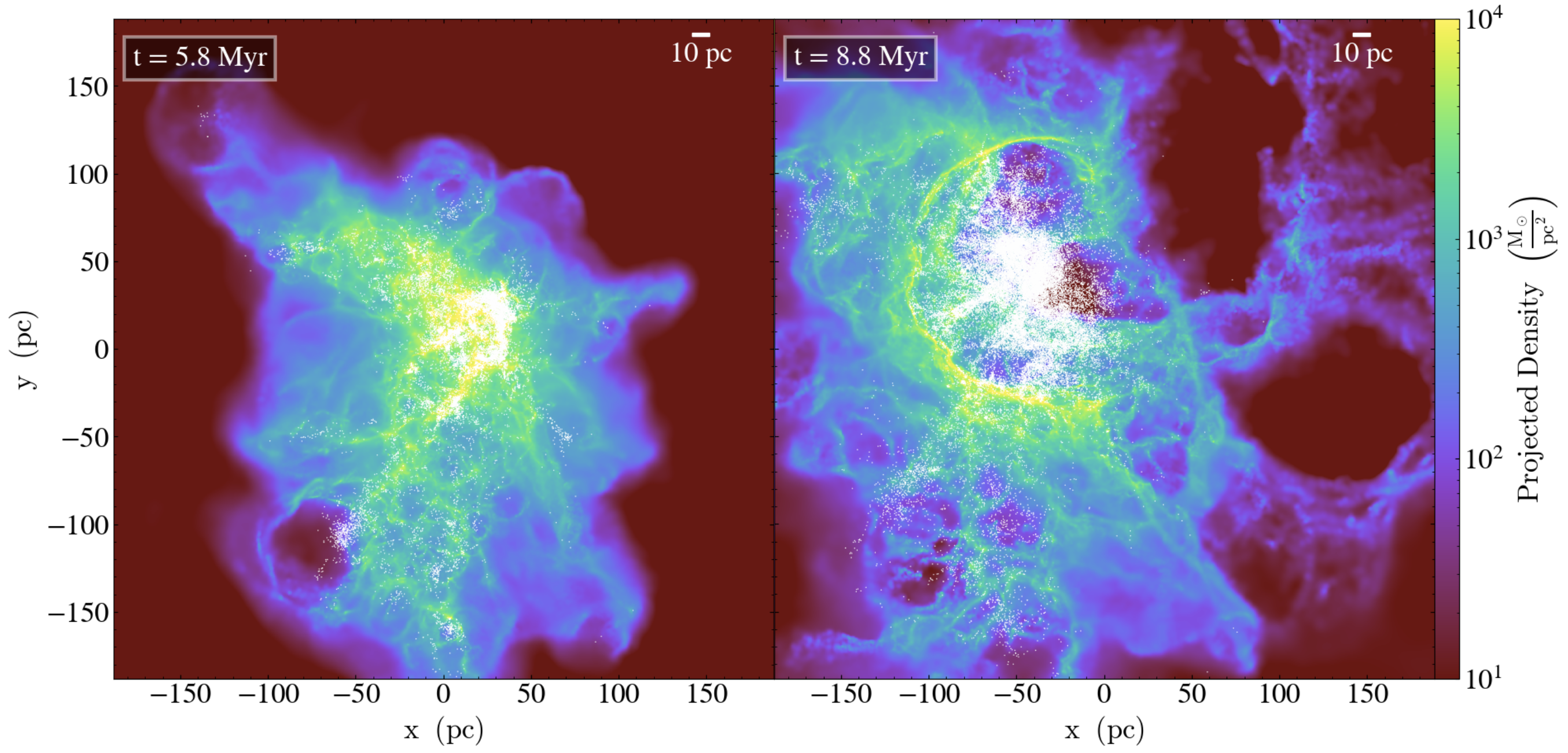}
    \caption{Snapshots from the simulation of \citet{Grudic2021_Associations} showing the formation of an arc structure similar to Barnard's Loop and the Orion Shell. The left snapshot was taken at a code time of 5.8 Myr, and the right snapshot was taken at a code time of 8.8 Myr. White dots show stars in the simulation, and the color bar refers to projected gas density. Roughly 20 supernovae exploded in the 3 Myr gap, creating an arc roughly 50 pc in radius. 
    \label{fig:GrudicSim}} 
\end{figure*}

As suggested in the timeline in Figure \ref{fig:orion_timeline}, we argue that Barnard's Loop was formed from the expelled gas in the region of OBP-B1. The gas was propelled by stellar feedback from OBP-B1 and the rest of the Orion Belt association, with the current structures finalized by a supernova from OBP-B1. Due to the proximity of OBP-B1 to Barnard's Loop, the effects from one or two supernovae from OBP-B1 may have played an outsize role in creating the arc-like structure visible today. Indeed, both Barnard's Loop and the Orion Shell possess radii of $\sim 50$ pc. To qualitatively illustrate this formation mechanism, we compare our observed data with molecular cloud simulations from \citet{Grudic2021_Associations}. 

\citet{Grudic2021_Associations} analyze the formation of bound and unbound clusters within molecular clouds. They simulate a number of clouds with different properties, but their fiducial model of $4 \times 10^{6}$ M$_{\odot}$ produces a structure that resembles Barnard's Loop exceptionally well in structure and size. Figure \ref{fig:GrudicSim} shows two snapshots from this simulation: one prior to any supernova explosions and one after roughly $\sim 20$ supernovae have gone off in the simulation. The latter snapshot features an arc with a radius of 50 parsecs, the same size as Barnard's Loop. Additionally, this structure is produced roughly 8 Myr after the first stars have formed, corresponding well with our timeline proposed in Section \ref{sec: timeline}. 

The simulation includes both stellar winds and radiation, which begin shortly after stellar birth (see \citet{Grudic2021_Associations} for more details of the stellar feedback prescription). Winds and radiation appear to have a minor effect on the gas after a few Myr in the simulation, having swept up only a few smaller shells by the time of the first snapshot shown in Figure \ref{fig:GrudicSim}. At this point, the first supernova explodes. Around 20 more supernovae explode over the next 3 Myr, and almost all of them occur near the dense center of the stellar mass found near the center of the resulting loop structure. The supernovae appear to act in concert, providing continuous momentum to the coherently expanding arc while blowing out less dense material away from it. The resulting arc structure is a superposition of many shells created by individual supernovae, first appearing around $t = 8.3$ Myr in the simulation and persisting for about 1 Myr.

Using this simulation as an analog for the Orion region, we posit that supernovae throughout the Orion region have also contributed to the formation of Barnard's Loop and the Orion Shell. However, supernovae produced close to the center of the arc in the simulation appear to dominate the resulting superposition of shells forming the arc. As a result, we argue that supernovae from OBP-B1 over the past few hundred thousand years to few million years likely played an outsize role in shaping the morphology of Barnard's Loop due to the cluster's proximity to the Loop's flux center. However, OBP-B1 is likely far from the only contributor to the formation of Barnard's Loop. As previously suggested in various works (i.e. \citet{Bally2008}), somewhere on the order of 10-20 supernovae are thought to have exploded in the Orion region. The gas dynamics in this simulation suggest that these supernovae might act in concert to produce both Barnard's Loop and the gas morphology throughout the Orion-Eridanus region. More evidence supporting this claim and constraining the timing and positioning of the supernovae will be presented in Paper II.

We note that this simulation is not a perfect analog for the Orion region. Barnard's Loop is visible in ionized gas and invisible in our 3D dust map, while the simulation produces an arc in dense gas. Additionally, this simulation is likely a factor of a few times more massive than the Orion region. Finally, the simulation assigns every star particle the same mass and computes feedback quantities in an IMF-averaged way. This likely would not affect average large-scale dynamical quantities too much, but it could introduce a number of artifacts in the simulation that change what specific structures are produced. Nevertheless, given the correspondence between the Orion Shell and Barnard's Loop, it remains a useful comparison tool for the Orion region.

\section{Conclusions}
\label{sec: conclusions}
\citetalias{Swiggum2021} identified a coherently expanding cluster of stars, OBP-B1, which we find is located in the center of the Orion Shell. OBP-B1 exhibits radial expansion out to foreground dust clouds, and radial expansion of gas in Orion also appears to emanate from a point close to OBP-B1. There is anisotropy in the expansion of OBP-B1, which can partly be explained by ``gravitational drag", or gravitational acceleration of the stars by gas clouds surrounding the cluster. Additional comparisons with simulations regarding gravitational drag are needed to fully quantify the effect.

OBP-B1 lies near a large enhancement of $^{26}$Al, a prominent tracer of supernovae. Using stellar kinematic data and verifying ages with stellar isochrones, we find that OBP-B1 likely produced one to a few supernovae within the past few Myr. Simple momentum analyses of the Orion Shell also suggest this is a reasonable estimate for the number of supernovae from the cluster. Coupled with early stellar feedback, supernovae from OBP-B1 could have supplied much of the momentum necessary to form the Orion Shell. 

The Orion Shell possesses the same radius as Barnard's Loop. The center of expansion of OBP-B1 falls close to the center of Barnard's Loop as well, suggesting that the expansion of OBP-B1 and Barnard's Loop are coupled. Analyzing two different simulations that serve as analogs for the Orion region, we contend that recent supernovae from OBP-B1 played a large role in shaping the current structure of Barnard's Loop. Additionally, OBP-B1 is likely responsible for the progenitor of the Barnard's Loop bubble identified in \citet{Ochsendorf2015}. 

Finally, all three Orion molecular clouds --- Orion A, B, and $\lambda$ --- and many regions of gas traced by \citet{Grosschedl2021} lie on the surface of the Orion Shell in 3D. Extrapolating the current velocities of gas in the Orion complex backwards in time reveals that much of the gas once resided close to the center of OBP-B1. Coherent expansion of both stars and gas from around the center of OBP-B1 implies that much of the present star formation in Orion is triggered by stellar feedback from OBP-B1 and the rest of the Orion OB1 association.

\vspace{5mm}

\acknowledgements
We are grateful to Roland Diehl for access to the map of $^{26}$Al. We would like to thank Bob Benjamin, Elena D'Onghia, Lars Hernquist, Charlie Conroy, and Charlie Lada for useful discussions about this work. 

The visualization, exploration, and interpretation of data presented in this work was made possible using the \texttt{glue} visualization software, supported under NSF grant numbers OAC-1739657 and CDS\&E:AAG-1908419. CZ acknowledges that support for this work was provided by NASA through the NASA Hubble Fellowship grant \#HST-HF2-51498.001 awarded by the Space Telescope Science Institute, which is operated by the Association of Universities for Research in Astronomy, Inc., for NASA, under contract NAS5-26555. JCF is supported by a Flatiron Research Fellowship through the Flatiron Institute, a division of the Simons Foundation. SB acknowledges support from the Center of Theory and Computations (CTC) at the University of Maryland, College Park. JA acknowledges support from the Data Science Research Center and the TURIS Research Platform of the University of Vienna. Support for MYG was provided by NASA through the NASA Hubble Fellowship grant \#HST-HF2-51479 awarded by the Space Telescope Science Institute, which is operated by the Association of Universities for Research in Astronomy, Inc., for NASA, under contract NAS5-26555. JG acknowledges funding by the Austrian Research Promotion Agency (FFG) under project number 873708.\\
\textit{Software:} \texttt{astropy} \citep{AstropyCode}, \texttt{glue} \citep{glueviz2017}, \texttt{numpy} \citep{Numpy2020}, \texttt{PyVista} \citep{PyVista2019}, \texttt{scipy} \citep{Scipy2020}

\bibliography{main}{}
\bibliographystyle{aasjournal}

\nocite{*}

\end{document}